 \def\CM{{\cal M}} \def\CL{{\cal L}}

\def\nn{\nonumber}        
\def\bm#1{\mbox{$\boldmath{#1}$}}
\def\prl#1{Phys.\ Rev.\ Lett.\ {\bf #1}}
\def\pr#1{Phys.\ Rev.\ {\bf #1}}
\def\np#1{Nucl.\ Phys.\ {\bf #1}}
\def\pl#1{Phys.\ Lett.\ {\bf #1}}
\def\prt#1{Phys.\ Reports.\ {\bf #1}}
\def\jpsj#1{J.\ Phys.\ Soc.\ Jpn.\ {\bf #1}}
\def\be{\begin{equation}}
\def\ee{\end{equation}}
\def\Be{\begin{eqnarray}}
\def\Ee{\end{eqnarray}}
\def\ba{\begin{array}}
\def\ea{\end{array}}
\documentstyle[aps,epsf,preprint]{revtex}
\tightenlines
\begin{document}
\draft

\title{An Isobaric Model for Kaon photoproduction}

\author{Bong Soo Han, Myung Ki Cheoun
\footnote{Corresponding author, e.mail : cheoun@phya.yonsei.ac.kr}, 
K.S.Kim, and Il-Tong Cheon}

\address{
Department of Physics, Yonsei University, Seoul, 120-749, Korea
\\ (2 December, 1999)}

\maketitle
\begin{abstract}
The  kaon photoproduction  is analyzed up 
to $E_\gamma^{\rm Lab}$=2.0 GeV 
by using an isobaric model based on effective Lagrangians and
by taking a cross symmetry into account.
Both {\it pseudovector}
and {\it pseudoscalar} couplings for kaon-baryon-baryon
(baryon spin=1/2)
interactions are considered with form factors.
A vector meson($K^*(890)$),
an axial vector meson($K_1(1270)$), nucleon resonances($J\le5/2$), and hyperon
resonances($J\le3/2$) are treated as participating particles.
By determining unknown coupling constants through 
a systematic fitting of the differential cross section,
the total cross section, the single polarization observable,
and the radiative kaon capture branching ratio to their experimental 
data, we find out a simple model which reproduces all the experimental
data well.
\end{abstract}
\vspace{1cm}

\section{Introduction}

It is believed that quantum chromodynamics(QCD) is a basic theory of the
strong interaction. The QCD was born by combining the Yang and Mills's
non-Abelian gauge theory with the quark model.
It has a $SU_L(N_f)\otimes SU_R(N_f)$ symmetry($N_f$ is the flavor number)
in a massless limit of quark and has a non-perturbative property in the
intermediate energy region. In the non-perturbative vacuum,
the $SU_L(N_f)\otimes SU_R(N_f)$  symmetry is spontaneously broken down to
$SU_V(N_f)$, and the massless Goldstone bosons, which are 
pseudoscalar mesons with spin and parity $J^P=0^-$, appear.
Because of the non-perturbative property of QCD in 
the intermediate energy region, we can not obtain sufficient information
about nuclear force by  directly tackling the QCD. Therefore, in the
region, we investigate the nuclear physics by an effective theory 
which have the basic symmetry in QCD. 
Since the pseudoscalar meson plays important roles in the intermediate
energy nuclear physics using such effective theory, 
understanding the property of the
pseudoscalar meson is an inevitable study in the nuclear physics at the
energy region. In specific, photo- and electro-production
of pseudoscalar mesons is complementary to the reaction, such as
the electron-nucleus scattering, $\pi$-nucleus scattering, muon capture, and
radiative pseudoscalar meson capture.
Compared with the pseudoscalar meson,  the real or virtual photon,
as a probing particle, is very weakly absorbed in the nucleus.
Therefore, the photo- or electro-production of pseudoscalar meson gives more
cleaner and more reliable information on the property of pseudoscalar 
meson interactions with nuclei than the hadron-induced reactions.

On the other hand, with increasing  interests in hypernuclei, 
the kaon electromagnetic productions on nuclei have
been interested as the reaction for producing the
hypernuclei. The study of
 kaon photoproduction (KP) on a nucleon started in the
late 50's, but a comprehensive description of the underlying
reaction mechanism is still not available because 
copious number of  nucleonic and hyperonic resonances may
intervene in the process 
due to the high threshold energy($E_\gamma^{\rm Lab}=910$MeV ) of 
the reaction even near threshold. Moreover,  most of the relevant 
coupling constants are still unknown.

However, with the advance of high energy and 
high duty cycle electron accelerators
and detectors at the Thomas Jefferson National Accelerator Facility(TJNAF),
ELectron Stretcher Accelerator (ELSA) and European Synchrotron Radiation
Facility(USRF), the high-current and polarized beams in the energy domain of a few
GeV's  are provided. Consequently, the sufficient and high-precision experimental
data of kaon photo- and electro-production are available now or will be in
the near future.
Therefore, it is an urgent task to establish and to improve the  theoretical
models about the kaon photo(electro)production.

Most theoretical studies for KP, so far, have been performed
by using dispersion relations\cite{Hats62}-\cite{Neli63},
multipole analysis\cite{Scho70}, quark-based models\cite{Kuma94}-\cite{Land95},
and  isobaric models\cite{Gour63}-\cite{Sagh97}.
In the dispersion theory or multipole analysis, amplitudes are obtained in
K-$\Lambda$ center of mass (c.m.) reference frame, 
so that transforming them into other
frames is cumbersome and ambiguous. In addition, it is difficult to discuss
non-local and off-shell effects, which turned out to play a significant role in
nuclear applications of the elementary KP amplitudes.
Moreover, in the dispersion theory, due to a high-energy threshold of
the reaction, the multi-pion channels
$\gamma + N \rightarrow N + m \pi$ with $m$=1 to 4 are already
open  for $K^+\Lambda$ process. 
The inclusion of these reactions leads to very complicated integral
equations among the partial amplitudes. To solve the integral equations,
unjustifiable approximations are needed.
Advantage of using a quark-based model is to describe the
reaction by an unified scheme and to explain the reaction well with relatively
less parameters than other models. But nuclear application of the model is
also not easy.

Finally, the isobaric models are widely used methods to investigate the
KP between threshold and roughly $E_\gamma^{\rm Lab}=2.0$ GeV region.
Based on pioneer works of Thom\cite{Thom66} in 1960's  and Renard and
Renard \cite{Rena71} in 1970's,  Hsiao and Cotanch \cite{Hsia83} and Adelseck, Bennhold, and
Wright\cite{Adel85}  revived  the models in 1980's.

All these formulas use the Feynman diagrammatic technique where
vertex functions are obtained from a relevant effective Lagrangian.
As well known, 
there are two coupling types of kaon and baryons with $J=\frac{1}{2}$ 
in strong vertex. One is a pseudoscalar(PS) coupling where
a kaon-nucleon-hyperon ($KNY$) vertex is described by $g_{KNY}\gamma_5$ with a
PS coupling constant of $KNY$, $g_{K N Y}$, 
and the other is a
pseudovector(PV) coupling in which the $KNY$ vertex is described by
$\frac{f_{KNY}}{M_N+M_Y} q\!\!\!/ \gamma_5$,
where $f_{KNY}$, $M_N,\ M_Y$, and $q$ are a PV-coupling constant,
a nucleon mass, a hyperon mass, and an outgoing
kaon four-momentum, respectively.

Through a chiral rotation\cite{Wein67},
the nonlinear $\sigma$ model, in which $\pi NN$ coupling is PV type,
is related to the linear sigma model adopting PS coupling.
In an infinite sigma mass limit, both models give identical
results. But it is valid only at a tree order because the nonlinear
version encounters divergences in one or more loops
calculations, so that one can not confirm the equivalence
between two coupling schemes as far as one includes higher order terms as 
an effective interaction under a tree order.

In kaon(or pion) photoproduction, without the Pauli-type
electromagnetic interactions stemming  from  meson loop(s) i.e. higher order 
terms,
two coupling schemes give the same predictions \cite{Cohe89}. 
Since the Pauli interaction gives large contribution
it is  unavoidable to consider the interaction
in the tree approximation calculation. The two coupling
schemes, consequently, give different predictions.

In case of a charged pion photoproduction, the deviation from
both coupling schemes turned out to be
small enough to neglect, because the Kroll-Rudermann term in PV scheme, which
is also included in the nucleon pole term in PS scheme, 
is dominant at the threshold.
For a neutral pion production,
the deviation is significant \cite{Fria77} because the Kroll-Rudermann term 
disappears in the PV description.
The PS coupling description overestimates the experimental
 ${E_{0+}}$ amplitude at the threshold about factor 10
even taking the neutral vector mesons' contributions into account,
while the PV scheme, at the threshold, satisfies a conventional low energy
theorem(LET)\cite{Baen70}.
That is the  reason why the PV scheme  has been  preferred
to the PS coupling in describing the pion photoproduction near threshold.
Although recent experiments \cite{Berg96,Krus95} and many theoretical
researches\cite{Naus90,Sche91,Drec92,Sche93} about $E_{0+}$ show 
that the LET disagreed to experimental values more or less, the
superior prediction of PV Born terms to that of PS Born terms remains
true at the threshold.
Beyond threshold two descriptions yield nearly
identical predictions
\cite{Lage77} because the contribution from $\Delta(1232)$ dominates over that of
other spin $\frac{1}{2}$ baryons.

In the KP, 
it is not still clear which is better  of the
 two coupling schemes. Moreover, the contributions of the spin $\frac{1}{2}$ particles
are comparable to those of particles with other spins.  Therefore,
the difference between the PV and PS
 coupling schemes in this reaction is expected to be 
much larger than in the pion photoproduction.

Nevertheless, until now,  most  calculations based on an isobaric model were
carried out by the PS coupling scheme. 
Some works\cite{Benn87,Cheo96,Feus98} were done by using the PV
coupling, but they used models oversimplified. Namely, 
Bennhold {\it et al.} \cite{Benn87} tried to
fit phenomenologically the
available data for $\gamma +p \rightarrow K^+ + \Lambda$ by using the model of
Adelseck {\it et al.}\cite{Adel85}
without $K_1$ axial vector meson, and obtained
 very different sets of coupling constants for
the two coupling schemes.
Feuster {\it et al.} \cite{Feus98}, by using a 
unitary model, investigated
 photon- and meson-induced reactions. In their calculation only nucleon
resonances with $J \leq \frac{3}{2}$  were considered with 
strong form factors in a gauge
invariant way by using the Haberzettl's and Ohta's gauge prescriptions
explained in Section \ref{secform}.  In the paper, they studied 
the difference between the PV- and PS-coupling schemes and between
the two gauge prescriptions.
However, they didn't consider some important contributions :
Firstly, the spin $\frac{5}{2}$ particle was excluded in their calculation.
It was shown to play an important role in reproducing 
the polarization data in Ref. \cite{Sagh96,Davi96}.
Secondly, they didn't take the contribution from hyperon
resonances into account. 
Their contributions play a vital role in this reaction. 
For example, in the previous works for the radiative kaon
capture\cite{John61,Velo69,Burk85,Work88,Jong92},
it was founded that $\Lambda(1405)$ gave a dominant contribution 
to the reaction. Since the radiative kaon capture and the
 KP are related to each other by a crossing symmetry
 they  should be
simultaneously  parameterized.
Therefore, 
in the more extended model spaces,
the more extensive study using the PV-coupling scheme and 
scrupulous comparison  for both
coupling schemes are needed.

One of main goal of this paper is
to construct the pseudovector model to reproduce well all of existing
 experiment data of the reaction and the radiative kaon capture
 up to $2.1$ GeV of the photon energy by using a $SU(3)$ symmetry and
 nonrelativistic quark model(NRQM), and to compare the PV-model and
 the PS-model, 
and to investigate the effect of off shell strong form factor
 in the reaction.

In the study up to 2.1 GeV of photon energy, we adopt an 
isobaric model based on the 
effective Lagrangian method and consider a $K^*$ vector
meson and a $K_1$ axial vector meson as exchanged resonance particles
in $t-$channel, the nucleon resonances with spin($J\leq$  $\frac{5}{2}$) in
$s-$channel, and the hyperon resonances with spin($J\leq$  $\frac{3}{2}$) in
$u-$channel. The unknown parameters are determined by a fitting
procedure, which is carried out by imposing the constraint
conditions from about 20 $\%$ 
broken $SU(3)$ symmetry and the NRQM\cite{Isgu77,Juri83} for 
strong and electro-magnetic decay widths of resonances. 
As for couplings, pseudovector (PV) coupling
Lagrangians and pseudoscalar (PS) Lagrangians are used for their
mutual comparison.

Since the hadron is treated as
a point particle, effect of the internal structure of the
hadron is not considered in the effective Lagrangian approach. 
It causes a divergent behavior of the reaction
amplitude  as increasing an incident photon energy. Therefore, we
introduce the form factor in the strong vertex. However, it
spoils a gauge invariance because 
the strong form factor is inserted into the vertex by hand. 
The hadronic form factor is, thus, introduced 
in a gauge invariant manner by using Haberzettl's and Ohta's prescriptions
and dependence of this reaction on the gauge prescriptions is examined in our 
model.
This reaction is also studied without the strong form factors for 
the comparison
with the case form factor used.

This paper is constructed as the following order.
In Section II, with a simple explanation of kinematics and invariant
amplitude used through this paper, 
definitions of the physical
observables such as differential cross section, single
polarization observables, and radiative capture are given.
In Section \ref{Clag}, interaction Lagrangians and
invariant amplitudes for each PV- and PS-coupling schemes are
presented. The form factors and gauge
prescriptions are described in  Section \ref{secform}.
In Section \ref{secfit}, we discuss our fitting strategy to determine
the coupling constants and to select the particle included in our model.
In Section
\ref{sec:result},  our results and discussions are given.
Finally, conclusions are presented in Section \ref{sec:con}.

\section{Basic Formalism}

\subsection{Kinematics}

In this section, we present kinematical relations among 
kinetic variables and
an invariant amplitude of the kaon photoproduction:
\begin{equation}
\gamma(k) + N(p) \longrightarrow K^+(q) + \Lambda(p')~,
\label{reac}
\end{equation}
where $p$, $k$, $ p'$ and $q$ are momenta of nucleon(N),  photon$(\gamma)$,
$\Lambda$, and kaon($K^+$), respectively.
The Mandelstam variables are given below
\begin{equation}
s = (p+k)^2=(q+p')^2, ~u = (q-p)^2= (k-p')^2, ~
t = (k-q)^2=(p'-p)^2
\label{man}
\end{equation}
with a well-known relation $
s+u+t =M^2+M_\Lambda^2+m_K^2$, 
where $M$, $m_K$, and $M_\Lambda$ are 
the masses of $N$, $K^+$, and $\Lambda$, respectively.

\begin{figure}
\begin{center}
\leavevmode
\epsfxsize=5.0in
\epsfysize=3.0in
\epsfbox{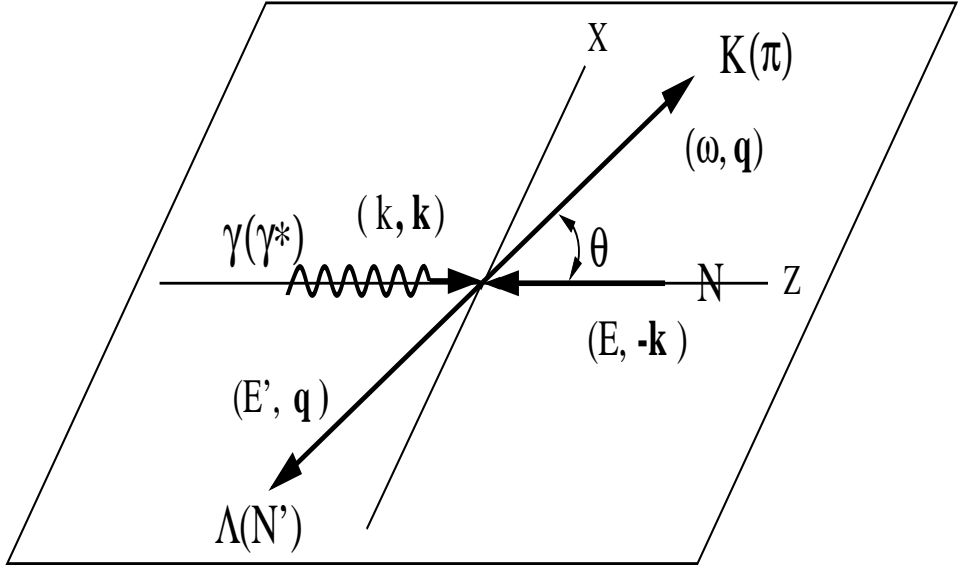}
\end{center}
\caption[The reaction plane of the the KP
(pion electroproduction)in $K-\Lambda$ center of mass reference frame.]{
The reaction plane of the KP in $K-\Lambda$ center
of mass reference frame.
We take the x-z plane as a reaction plane.
An incident real photon goes along 
the $+Z$-axis and $\theta$ is a production angle
 of the kaon.}
 \label{fig:re-plane}
\end{figure}

As depicted in Fig. \ref{fig:re-plane}, we choose 
the $+z$-axis  along the incident photon
direction and  $\theta$ as the production angle between
the photon and the kaon. The 
momenta of the initial and the final particles are
thus expressed in the c.m. as
\Be
k^\mu &=& (k^*,\ 0,\ 0,\ k^*), p^\mu = (E,\ 0,\ 0,\ -k^* )~,~ \nn \\
q^\mu &=& (\omega,\ q^* \cos\theta,\ 0,\ q^* \sin\theta )~,~ 
p'^\mu = (E',\ -q^* \cos\theta,\ 0,\ -q^* \sin\theta )~,
\Ee
where $k^*$ and $q^*$ are the magnitude in the c.m. system 
given in terms of photon lab. energy $k_{lab}$
\begin{equation}
k^* = \frac{M}{W} k_{lab}~,~ q^* = \frac{\sqrt{ 
(s-M^2-m_K^2)^2-4 m_K^2 M^2}}{2 W}~,
\end{equation}
where $W=\sqrt{s}$.

The ${\cal S}$-matrix is expressed as follows
\begin{equation}
{\cal S}_{fi} =\frac{1}{(2\pi)^2} \delta^4 (p+k-p'-q)
    \left[ \frac{M_p M_\Lambda}{4 E_\Lambda E_p E_\gamma E_K} \right]^{1/2} {\cal M}_{fi}~.
\label{sfi}
\end{equation}
To keep the gauge invariance of ${\cal M}_{fi}$ the
following self gauge
invariant operators  ${\cal O}_j$ have been used \cite{Adel85}
\begin{eqnarray}
{\cal O}_1 &=& \frac{1}{2}\gamma_5 [k\!\!\!/, \epsilon\!\!\!/], \nonumber \\
{\cal O}_2 &=& 2 \gamma_5 (\epsilon \cdot p k \cdot p' -\epsilon\cdot p' k\cdot p ), \nonumber \\
{\cal O}_3 &=&  \gamma_5 ( \epsilon\!\!\!/ k\cdot p -k\!\!\!/ \epsilon \cdot p), \nonumber \\
{\cal O}_4 &=&  \gamma_5 ( \epsilon\!\!\!/ k\cdot p' -k\!\!\!/ \epsilon \cdot p').
\label{oj}
\end{eqnarray}
The ${\cal M}_{fi}$ is expanded, in terms of ${\cal O}_j$, as
\begin{equation}
\CM_{fi}=\bar{u}_\Lambda (p',s') \sum_{j=1}^{4} {\cal A}_j {\cal O}_j 
u_N (p,s)~,
\label{mfi}
\end{equation}
where  $u_N (p,s)$ and $\bar{u}_\Lambda (p',s')$ are Dirac-spinors of proton and lambda
and  $s$ and  $s'$ represent their spin states. 
We write also ${\cal M}_{fi}$ in terms of
Chew, Goldberger, Low, and  Nambu(CGLN) amplitudes ${\cal F}_i$\cite{CGLN57}
\begin{equation}
{\cal M}_{fi}^\lambda=\biggl(\frac{\sqrt{M_\Lambda M}}{4 \pi W} \biggr)^{-1}
   \chi_f^\dagger{\cal F}^\lambda \chi_i~,
\label{eqMF}
\end{equation}
with
\begin{equation}
{\cal F}^\lambda={\cal F}_1 \vec{\sigma}\cdot\vec{\epsilon}^\lambda +
{\cal F}_2 i \vec{\sigma}\cdot\hat{q}
\vec{\sigma}\cdot(\hat{\epsilon}^\lambda\times\hat{k})+
{\cal F}_3 \vec{\sigma}\cdot\hat{k}\hat{q}\cdot\hat{\epsilon}^\lambda+
{\cal F}_4 \vec{\sigma}\cdot\hat{q} \vec{\epsilon}^\lambda\cdot\hat{q}~,
\end{equation}
where $\lambda$ represents a polarization state of the photon.
In connection with the ${\cal A}_i$,  the  ${\cal F}_j$ are given  by
\begin{eqnarray}
{\cal F}_1 &=& \frac{|\vec{k}|}{4 \pi}
\sqrt{\frac{E_\Lambda +M_\Lambda}{2 W}}\left[ {\cal A}_1-\frac{W+M}{2}{\cal A}_3
 -\frac{k.p'}{W-M}{\cal A}_4 \right], \nonumber \\
{\cal F}_2 &=& -\frac{|\vec{k}|}{4 \pi}
\sqrt{\frac{E_\Lambda -M_\Lambda}{2 W}}\left[ {\cal A}_1+\frac{W-M}{2}{\cal A}_3
+\frac{k.p'}{W+M}{\cal A}_4 \right], \nonumber \\
{\cal F}_3 &=& -\frac{|\vec{k}| |\vec{q}|}{4 \pi}
\sqrt{\frac{E_\Lambda +M_\Lambda}{2 W}}\left[ (W-M){\cal A}_2-
 {\cal A}_4 \right], \nonumber \\
{\cal F}_4 &=& \frac{|\vec{k}|}{4 \pi}
\sqrt{\frac{E_\Lambda -M_\Lambda}{2 W}}\left[ (W+M){\cal A}_2+
 {\cal A}_4 \right]~.
\end{eqnarray}

\subsection{Differential Cross Section and  Single  Polarization Observables}
\label{secobs}

Using the ${\cal M}_{fi}^\lambda$ in Eq.(\ref{eqMF})
we can write a differential cross section  in the c.m. frame
of $K-\Lambda$ as follows
\begin{equation}
\left( \frac{d\sigma}{d\Omega} \right)_{\mbox{c.m.}}=
        \frac{1}{4} \sum_{s_i,s_f,\lambda} \frac{q^*}{k^*}
        \left|\frac{\sqrt{M_\Lambda M_p}}{4 \pi W} \CM_{fi}^\lambda \right|^2.
\label{eqds}
\end{equation}
The  $\Lambda$-polarization asymmetry($P$),
beam polarization symmetry($\Sigma$), and target-polarization asymmetry($T$)
are defined by\cite{Adel90}
\begin{eqnarray}
P&=&\frac{(d\sigma/d\Omega)^{(+)}-(d\sigma/d\Omega)^{(-)}}{(d\sigma/d\Omega)^{(+)}+(d\sigma/d\Omega)^{(-)}},     \nn \\
\Sigma&=&\frac{(d\sigma/d\Omega)^{(+)}-(d\sigma/d\Omega)^{(-)}}{(d\sigma/d\Omega)^{(+)}+(d\sigma/d\Omega)^{(-)}},\nn \\
T&=&\frac{(d\sigma/d\Omega)^{(\perp)}-(d\sigma/d\Omega)^{(\parallel)}}{(d\sigma/d\Omega)^{(\perp)}+(d\sigma/d\Omega)^{(\parallel)}}~,
\label{eqpol}
\end{eqnarray}
where +(--) represents that a proton(for $T$) or a lambda(for $P$)
is polarized parallel(antiparallel) to the
direction($\frac{\vec{k}^*\times\vec{q}^*}{|\vec{k}^*\times\vec{q}^*|}$).
$\perp(\parallel)$ denotes that the photon is linearly polarized
perpendicular(parallel) to the reaction plane.

\subsection{Branching Ratio of Radiative Kaon Capture}
\label{seckcp}

The KP is restricted by a crossing symmetry \cite{Pilk67}:
{\it The matrix element for a process containing an antiparticle of 4-momentum
$p_\mu$ in the initial (final) state is identical with the matrix element for
the ''crossed'' process , which contains the corresponding particle of 4-momentum
$-p_\mu$ in the final (initial) state}.
The ''crossed'' process of KP is the radiative kaon capture(RKC).
Therefore, we can easily obtain the RKC amplitude, $\CM_{rad}$, from the
KP
amplitude, $\CM_{pho}$, as follows
\Be
\CM_{rad}&\equiv&\CM_{K^-(q) + p(p) \longrightarrow \gamma(k) +\Lambda(p')}, \nn \\
&=&\CM_{pho}\equiv\CM_{\gamma(-k) + p(p) \longrightarrow K^+(-q) +\Lambda(p')}
~,
\Ee
from which we get the following changing rules among the Mandelstam 
variables 
\Be
&&  {\rm rad.\ cap.}\ \ \ \ \ \ \ \ \ {\rm pho.\ kaon} \nn \\
s&=&(p+q)^2 \longrightarrow (p-q)^2=u, \nn \\
u&=&(p-k)^2 \longrightarrow (p+k)^2=s, \nn \\
t&=&(k-q)^2 \longrightarrow (q-k)^2=t.
\Ee
Therefore, the amplitude $\CM_{rad}$ is given by\cite{Ji88}
\be
\CM_{rad}(s,u,t)=\CM_{pho}(u,s,t).
\label{rad-amp}
\ee

The only available data for the radiative kaon capture is a 
branching ratio which is defined by
\be
R_{\gamma \Lambda} =\frac{\Gamma (K^- p \rightarrow \gamma \Lambda )}{\Gamma (K^- p \rightarrow \rm{all})}.
\ee
In a kaonic hydrogen atom, the kaon, whose momentum is approximately zero, is
strongly captured from an S-state. Since a range of strong interaction is
very short the kaon wave function is approximated as the value 
evaluated at the proton i.e. $\phi_K (0)$.
The decay rate for the process is thus given by
\be
\Gamma_{K^- p \rightarrow \Lambda \gamma}=|\phi_K(0)|^2
\frac{M_\Lambda k}{4 \pi (m_K +M) m_K} \frac{1}{2}
\sum_{{\rm \lambda, s_\Lambda,s_p}}|\CM_{rad}|^2~,
\label{radeq1}
\ee
where $\lambda,\ s_\Lambda$, and $s_p$ are a photon polarization, a $\Lambda$
spin, and a proton spin, respectively.
The total rate for all $K^- p$ processes is
\be
\Gamma_{K^- p \rightarrow {\rm all}} =2 W_p |\phi_K(0)|^2~,
\label{radeq2}
\ee
where $W_p$ is an imaginary part of the $K^- p$ pseudopotential and uses a 
value\cite{Burk85} $560\pm135$MeV fm$^3$.
By using Eqs. (\ref{radeq1}) and (\ref{radeq2}) we obtain 
the following branching ratio of this process
\be
R_{\gamma \Lambda}=\frac{M_\Lambda E_\gamma}{8 \pi W_p (m_K +M) m_K }\frac{1}{2}
\sum_{{\rm \lambda, s_\Lambda,s_p}}|\CM_{rad}|^2.
\ee

\section{Interaction Lagrangians and Invariant Amplitudes}
\label{Clag}

Under a tree approximation, the KP amplitudes on a proton
are obtained from the Feynman diagrams in Fig. \ref{fig:kp2}.
For convenience, we explicitly write down each reaction matrix ${\cal M}$
corresponding to each intermediate particle in Born terms, 
vector meson terms,
nucleon resonance terms($N^*$), and hyperon resonance terms($Y^*$) 
listed in
Table \ref{nrlist1}.
The total reaction matrix is thus obtained as a 
sum of  all terms mentioned above:
\begin{equation}
{\cal M}={\cal M}(Born)+\sum_{B=N^*,Y^*,K^*,K_1} {\cal M}(B).
\end{equation}
For more compact forms of Lagrangians, the following 
isospinors for $K, K^*, K_1$ are used
\begin{equation}
K=\left(\begin{array}{c}
            K^+ \nonumber \\
            K^0
        \end{array}
  \right)
\end{equation}
\begin{equation}
\begin{array}{cc}
K^*=\left(\begin{array}{c}
            K^{*+} \nonumber \\
            K^{*0}
        \end{array}
  \right),&
\; \;\;\;

K_1=\left(\begin{array}{c}
            K_1^+ \nonumber \\
            K_1^0
        \end{array}
  \right)~.
\end{array}
\end{equation}
For nucleon($N$) and nucleon resonances ($N^*$) we also write the isospinors
\be
\ba{cc}
N=\left(\ba{c}
         p \\
         n
        \ea \right),
&
\;\;\;\; N^*=\left(\ba{c}
         N^{*+} \\
         N^{*0}
        \ea \right)
\ea
\ee
and those for isovector $\vec{\Sigma}$ and its resonance $\vec{\Sigma}^*$ 
are given by
\be
\ba{cc}
 \vec{\Sigma}=\left(
    \ba{c}
        \Sigma^+ \\
        \Sigma^0 \\
        \Sigma^-
    \ea         \right),& \;\;\;\;
\vec{\Sigma}^*=\left(
    \ba{c}
        \Sigma^{*+} \\
        \Sigma^{*0} \\
        \Sigma^{*-}
    \ea         \right).
\ea
\ee

\subsection{Born Terms}
\label{sec:born}

Electromagnetic interaction Lagrangians for Born terms are given as
\begin{eqnarray}
\CL_{\gamma NN}&=&-e \bar{N}\left\{  \frac{1}{2}(1 +\tau_3  ) A\!\!\!/
      +\frac{1}{2} ((\kappa_p+\kappa_n)+(\kappa_p-\kappa_n)\tau_3)\frac{F^{\mu\nu}}{4 M}  \sigma_{\mu\nu} \right\}N, \nn \\
\CL_{\gamma \Lambda\Lambda}&=&-e \kappa_\Lambda \frac{F^{\mu\nu}}{4 M} \bar{\Lambda} \sigma_{\mu\nu} \Lambda,  \nn \\
\CL_{\gamma \Lambda\Sigma}&=&-e \kappa(\Sigma\Lambda)\left\{  \frac{F^{\mu\nu}}{4 M}(\bar{\Sigma}^3
        \sigma_{\mu\nu} \Lambda + \bar{\Lambda} \sigma_{\mu\nu} \Sigma^3 )\right\},  \nonumber \\
\CL_{\gamma KK}&=&ie (\partial_\mu K^+K^--\partial_\mu K^- K^+)
A^\mu,
\end{eqnarray}
where  $e$ is a charge of proton, $\kappa_p$, $\kappa_n$, and $\kappa_\Lambda$ are, respectively, 
the anomalous magnetic moment of a proton, a neutron, and $\Lambda$.
The
$\kappa(\Sigma\Lambda)$ is a $\Sigma^0 \Lambda$-transition magnetic moment.
Photon field strength is given as
\begin{equation}
F_{\mu\nu}=( \partial_\mu A_\nu - \partial_\nu A_\mu),
\end{equation}
and charged eigen-kaon fields $K^\pm$ are defined by
\be
K^\pm =\frac{1}{\sqrt{2}}(K^1 \pm i K^2)~.
\ee
Lagrangians in strong interactions are constructed by two coupling schemes,
PV and PS, which are expressed as
\begin{eqnarray}
\CL_{KN\Lambda}^{PV}&=&\frac{f_{KN\Lambda}}{m_K} (\partial^\mu K^\dagger  \bar{\Lambda} \gamma_\mu \gamma_5 N
       + \bar{N} \gamma_\mu \gamma_5 \Lambda \partial^\mu K), \nn\\
\CL_{KN\Sigma}^{PV}&=&\frac{f_{KN\Sigma}}{m_K}  (\partial^\mu K^\dagger\bar{\vec{\Sigma}}\cdot\vec{\tau} \gamma_\mu \gamma_5 N
      + \bar{N} \gamma_\mu \gamma_5 \bm{\tau}\cdot\vec{\Sigma} \partial^\mu
      K),\nn \\
\CL_{KN\Lambda}^{PS}&=&-i g_{KN\Lambda} (\bar{\Lambda} \gamma_5 N
       + \bar{N} \gamma_5 \Lambda ) K, \nn\\
\CL_{KN\Sigma}^{PS}&=& -i g_{KN\Sigma} (K^\dagger\bar{\vec{\Sigma}}\cdot\vec{\tau} \gamma_5 N
+ \bar{N} \gamma_5 \bm{\tau}\cdot\vec{\Sigma}  K),
\end{eqnarray}
where $m_K$ is a kaon mass, and  $f_{KN\Lambda}$, $f_{KN\Sigma}$, 
$g_{KN\Lambda}$, and $g_{KN\Sigma}$  are PV and PS 
coupling constants, respectively.

\begin{figure}
\begin{center}
\leavevmode
\epsfxsize=5.0inch
\epsfysize=5.0inch
\epsfbox{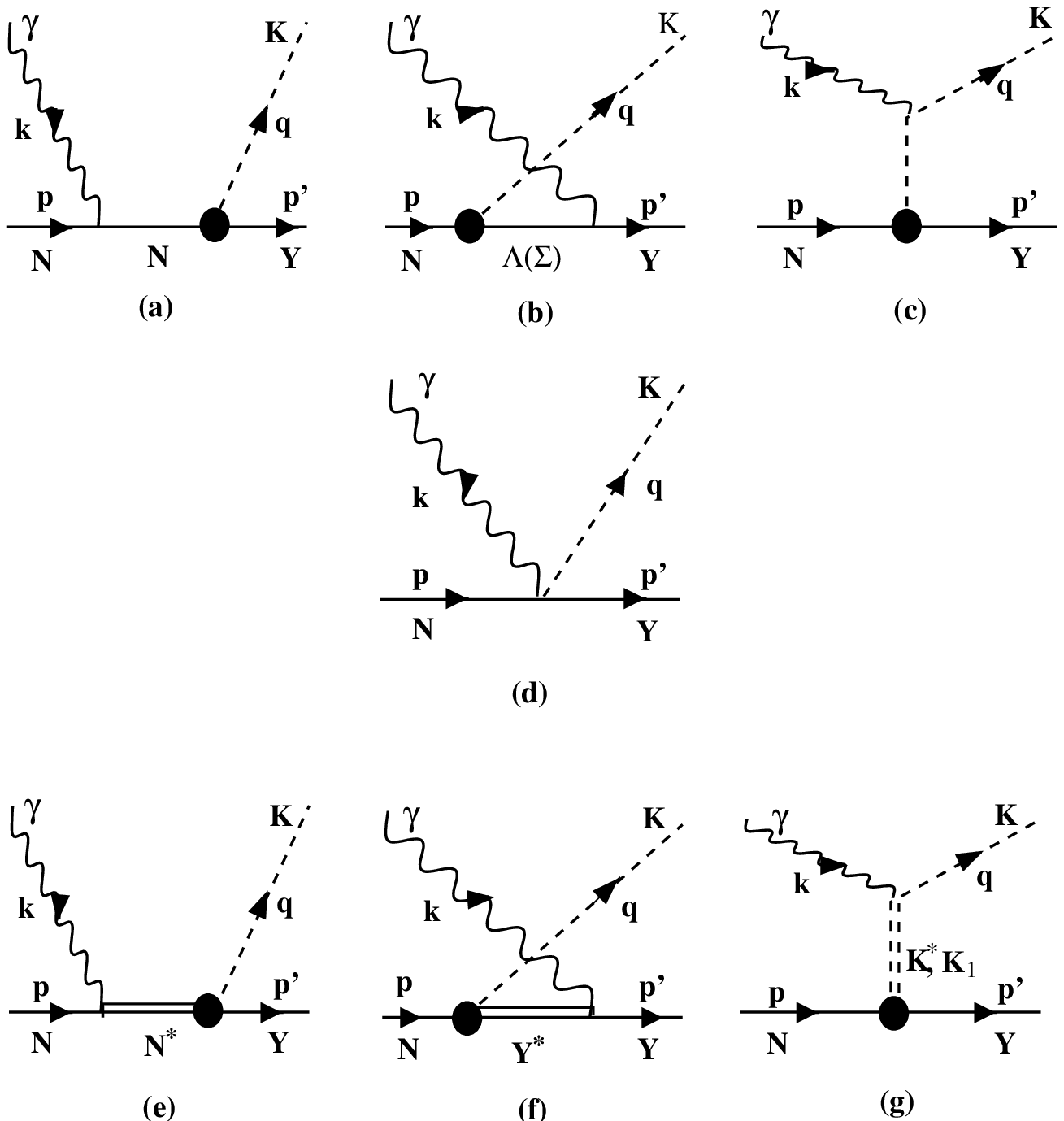}
\end{center}
\caption[Feynman diagrams of the KP]{
((a)-(c)) and ((e)-(g)) are diagrams of s-, u-, and t-channel
in Born terms and in resonance terms, respectively. The diagram (d) is the
Kroll-Rudermann diagram which is included only in the pseudovector coupling
scheme. Blobs in strong vertices denote hadronic form factors.
The N*(Y*) in intermediate states represents
the nucleon(hyperon) resonance.}
\label{fig:kp2}
\end{figure}

Using the vertex factor and the propagator calculated from the above
Lagrangians, the Born amplitudes for the PV coupling scheme are
written as
\begin{eqnarray}
{\cal M}_s^{PV}&=&{\cal M}_s^{PS} + e \frac{g_{{}_{KN\Lambda}}}{M+M_\Lambda}
 \bar{u}_\Lambda \gamma_5 \left(\epsilon \!\!/ + \mu_p \frac{1}{2} [k\!\!\!/ ,\epsilon \!\!/ ] \right) u, \nonumber \\
{\cal M}_{u(\Lambda)}^{PV}&=&{\cal M}_{u(\Lambda)}^{PS}+
e \mu_\Lambda \frac{g_{{}_{KN\Lambda}}}{M+M_\Lambda}
 \bar{u}_\Lambda   \frac{1}{2} [k\!\!\!/ ,\epsilon \!\!/ ] \gamma_5 u, \nonumber \\
{\cal M}_{u(\Sigma)}^{PV}&=&{\cal M}_{u(\Sigma)}^{PS}+
e  \mu(\Sigma\Lambda) \frac{g_{{}_{KN\Sigma}}}{M+M_\Sigma}
 \bar{u}_\Lambda   \frac{1}{2} [k\!\!\!/ ,\epsilon \!\!/ ] \gamma_5 u, \nonumber \\
{\cal M}_{KR}^{PV} &=& -e \frac{g_{{}_{KN\Lambda}}}{m_K} \bar{u}_\Lambda \gamma_5 \epsilon \!\!/ u, \nonumber \\
{\cal M}_t^{PV} &=& {\cal M}_t^{PS}=e g_{{}_{KN\Lambda}} \bar{u}_\Lambda \gamma_5 u \frac{\epsilon \cdot (2 q-k)}{t-m_K^2}~,
\end{eqnarray}
where $\mu_B = { {\kappa_B} \over { 2M}}$ for $ B = p, \Lambda, 
( \Sigma \Lambda)
$ and subscripts $s,\ u$ and $t$ refer to $s-,\ u-$ and
$t-$channel (diagrams (a)-(c) in Fig. \ref{fig:kp2}), respectively. The ${\cal M}_{KR}^{PV}$ is
an amplitude corresponding to the diagram (d) in Fig. \ref{fig:kp2}. The PS
coupling versions are given by
\begin{eqnarray}
{\cal M}_s^{PS} &=& e g_{{}_{KN\Lambda}}
  \bar{u}_\Lambda \gamma_5 \, \frac{p\!\!\!/ +k\!\!\!/ +M}{s-M^2}
\left(  \epsilon\!\!/ +  \mu_p \frac{1}{2} [k\!\!\!/ ,\epsilon \!\!/ ] \right) u, \nonumber \\
{\cal M}_{u(\Lambda)}^{PS} &=& e \mu_\Lambda g_{{}_{KN\Lambda}}
    \bar{u}_\Lambda  \frac{1}{2} [k\!\!\!/ ,\epsilon \!\!/ ]
\frac{p\!\!\!/ - q\!\!\!/ + M_\Lambda}{u-M_\Lambda^2} \gamma_5 u, \nonumber \\
{\cal M}_{u(\Sigma)}^{PS} &=& e \mu(\Sigma\Lambda) g_{{}_{KN\Sigma}}  \bar{u}_\Lambda  \frac{1}{2}  [k\!\!\!/ ,\epsilon \!\!/ ] \,
\frac{p\!\!\!/ - q\!\!\!/ + M_\Sigma}{u-M_\Sigma^2} \gamma_5 u~, 
\end{eqnarray}
with a relation $g_{{}_{KN\Lambda(\Sigma)}}= 
f_{{}_{KN\Lambda(\Sigma)}} \frac{M+M_{\Lambda(\Sigma)}}{m_K}$.

\subsection{Spin 1 Mesons}
\label{sec:spin1}

For the $K^*(890)(J^P=1^-)$, the following Lagrangians are used
\begin{eqnarray}
{\cal L}_{K^* N\Lambda} &=& \bar{\Lambda}\left\{ \biggl(G_{\rm v} \gamma_\mu +\frac{G_{\rm t}}{M+M_\Lambda}
\sigma_{\nu\mu}\partial^\nu \biggr) K^{*\mu\dagger} \right\}N+ h.c., \nonumber \\
{\cal L}_{\mbox{em}} &=& \frac{g_{\gamma KK^*}}{m}\epsilon_{\alpha\beta\mu\nu} (\partial^\alpha A^\beta)
(\partial^\mu K^\dagger )K^{*\nu} +h.c.,
\label{eqKv}
\end{eqnarray}
where $G_{\rm v}$ and $G_{\rm t}$ are, respectively, 
a strong vector coupling constant and
a tensor coupling constant, and $m$ is  an arbitrary mass parameter for making
the coupling constant $g_{\gamma KK^*}$ dimensionless.
We fix $m=1.0$ GeV as used in Ref. \cite{Adel90}.
The above Lagrangians give the following gauge- and 
Lorentz-invariant reaction matrix ${\cal M}$ corresponding to a diagram (g)
in Fig. \ref{fig:kp2}
\begin{eqnarray}
\CM &=&-\frac{1}{m} \epsilon_{\alpha\beta\tau\sigma}
k^\alpha\epsilon^\beta q^\tau \biggl( \frac{-g^{\sigma\mu}+
q'^{\sigma}q'^{\mu}/M_{K^*}^2}{t-M_{K^*}^2+i \Gamma M_{K^*}}\biggr)\nn \\
&& \times \bar{u}_\Lambda \biggl( G_V^{K^*}\gamma_\mu +
   i\frac{G_T^{K^*}}{M+M_\Lambda} \sigma_{\nu\mu} q'^{\nu} \biggr)u
\end{eqnarray}
with $G_V^{K^*}= g_{\gamma K^* K}*G_{\rm v},~ 
G_T^{K^*}= g_{\gamma K^* K}*G_{\rm t}
$.
Here the following identity is used 
\be
i \epsilon_{\alpha\beta\tau\sigma}\gamma_5\gamma_\sigma=
\gamma_\alpha\gamma_\beta\gamma_\tau- g_{\alpha\beta}\gamma_\tau
    -g_{\beta\tau}\gamma_\alpha +g_{\alpha\tau}\gamma_\beta ~.
\ee
The interaction Lagrangians for the axial vector meson $K_1(1270)(J^P=1^+)$ are taken by
\begin{eqnarray}
{\cal L}_{K_1 N\Lambda} &=&  \bar{\Lambda}\biggl( G1_{\rm v} \, \gamma_{\mu}  +
              \frac{G1_{\rm t}}{M+M_\Lambda}
\sigma_{\nu\mu}\partial^\nu \biggr) K_1^{\mu\dagger} \gamma_5 N + h.c., \nonumber \\
{\cal L}_{em} &=& -i\frac{g_{\gamma K_1 K}}{m} K^\dagger (\partial_\mu A_\nu \partial^\mu K_1^\nu -
                       \partial_\mu A_\nu \partial^\nu K_1^\mu ) + h.c. ~,
\end{eqnarray}
where $G1_{\rm v}$ and $G1_{\rm t}$ are a strong vector 
coupling and a tensor coupling constant, respectively. The following 
reaction matrix is then obtained
\begin{eqnarray}
{\cal M}&=&-\frac{1}{m}(\epsilon\cdot q' k_\mu-k\cdot q' \epsilon_\mu)
\biggl(\frac{-g^{\mu\nu}+q'^{\mu}q'^\nu/M_{K_1}^2}{t-{M_{K_1}^2}+i \Gamma M_{K_1}}\biggr) \nonumber \\
& & \times \bar{u}_\Lambda \left(G_V^{K_1} \gamma_\nu +i \frac{G_T^{K_1}}{M+M_\Lambda}
                    \sigma_{\alpha\nu} q'^\alpha\right)\gamma_5 u,
\end{eqnarray}
where coupling constants are defined by
$
G_V^{K_1}= g_{\gamma K_1 K}*G1_{\rm v},~
G_T^{K_1}= g_{\gamma K_1 K}*G1_{\rm t}
$.

\subsection{Spin-1/2 Isobar Terms}

Kaon interacts with spin-1/2 particles in two different ways, PS and PV 
types. We explicitly show PV type Lagrangians
and present PV amplitude in terms of a well known PS amplitude.

\subsubsection{Nucleon Resonances}
\label{sec:nucl1}

For the $S_{11}$ isobar, we take the Lagrangians:
\begin{eqnarray}
\CL_{\gamma N_{S_{11}} N} & =&-e \frac{\kappa(N_{S_{11}} N)}{4M} \biggl[\bar{N}_{S_{11}} \sigma_{\mu\nu}\gamma_5 N 
+\bar{N}\sigma_{\mu\nu}\gamma_5 N_{S_{11}} \biggr] F^{\mu\nu}, \nn \\
\CL_{K\Lambda N_{S_{11}}}^{PV}&=&\frac{f_{K\Lambda N_{S_{11}}}}{m_k}( \partial^\mu K^\dagger \bar{\Lambda} \gamma_\mu  N_{S_{11}}
       + \bar{N}_{S_{11}} \gamma_\mu  \Lambda \partial^\mu K),
\end{eqnarray}
where the field $N_{S_{11}}$ denotes a $S_{11}$ field. For this resonance we obtain
\begin{eqnarray}
{\cal M}^{PV} &=& {\cal M}^{PS} +e \frac{G_{S_{11}}}{2 M(M_{S_{11}}-M_\Lambda)}
                \frac{s-M_{S_{11}}^2}{s-M_{S_{11}}^2+i \Gamma M_{S_{11}}}
 \bar{u}_\Lambda
  \frac{1}{2} [k\!\!\!/ ,\epsilon\!\!\!/] \gamma_5 u~, \nn \\
{\cal M}^{PS} &=& -e \frac{G_{S_{11}}}{2 M} \bar{u}_\Lambda
  \frac{p\!\!\!/ +k\!\!\!/ +M_{S_{11}}}{s-M_{S_{11}}^2+i \Gamma M_{S_{11}}}
\frac{1}{2} [k\!\!\!/ ,\epsilon\!\!\!/] \gamma_5 u~,
\end{eqnarray}
where the coupling constant is
$
G_{S_{11}}=\kappa(N_{S_{11}}N)*g_{KN_{S_{11}} \Lambda},~ 
g_{KN_{S_{11}} \Lambda}= \frac{M_{S_{11}}-M_\Lambda}
{m_K}*f_{K\Lambda N_{S_{11}}}.
$

For $P_{11}$-nucleon resonance, Lagrangians are written as
\begin{eqnarray}
\CL_{\gamma N_{P_{11}} N} & =&-e \frac{\kappa(N_{P_{11}} N)}{4M}
\biggl[\bar{N}_{P_{11}} \sigma_{\mu\nu} N
+\bar{N}\sigma_{\mu\nu} N_{P_{11}} \biggr] F^{\mu\nu}, \nn \\
\CL_{K\Lambda N_{P_{11}}}^{PV}&=&\frac{f_{K\Lambda N_{P_{11}}}}{m_k}
 ( \partial^\mu K^\dagger \bar{\Lambda} \gamma_\mu \gamma_5  N_{P_{11}}
       +\bar{N}_{P_{11}} \gamma_\mu \gamma_5 \Lambda \partial^\mu K),
\end{eqnarray}
where the field $N_{P_{11}}$ denotes a $P_{11}$ field. For this resonance we get
\begin{eqnarray}
{\cal M}^{PV} &=& {\cal M}^{PS} +e \frac{G_{P_{11}}}{2 M(M_{P_{11}}+M_\Lambda)}
                \frac{s-M_{P_{11}}^2}{s-M_{P_{11}}^2+i \Gamma M_{P_{11}}}
 \bar{u}_\Lambda \gamma_5
  \frac{1}{2} [k\!\!\!/ ,\epsilon\!\!\!/]  u, \nn \\
{\cal M}^{PS} &=& e \frac{G_{P_{11}}}{2 M} \bar{u}_\Lambda \gamma_5
  \frac{p\!\!\!/ +k\!\!\!/ +M_{P_{11}}}{s-M_{P_{11}}^2+i \Gamma M_{P_{11}}}
\frac{1}{2} [k\!\!\!/ ,\epsilon\!\!\!/] u,
\end{eqnarray}
where the coupling constant is
$
G_{{P_{11}}}=\kappa(N_{P_{11}}N)*g_{KN_{P_{11}} \Lambda}, ~
g_{KN_{P_{11}} \Lambda}= \frac{M_{P_{11}}+M_\Lambda}{m_K}*
f_{K\Lambda N_{P_{11}}}
$, and $\kappa$ and $f$ are electromagnetic 
and strong coupling constants, respectively.

\subsubsection{Hyperon resonances}
\label{sec:hy1}

For $\Lambda^*(L_{I 2J}=S_{01})$ isobar, interaction Lagrangians are given by
\begin{eqnarray}
\CL_{\gamma \Lambda^*_{S_{01}} \Lambda} & =&-e \frac{\kappa(\Lambda^*_{S_{01}} N)}{4M}
    \biggl[\bar{\Lambda}^*_{S_{01}} \sigma_{\mu\nu}\gamma_5  \Lambda
+\bar{\Lambda}\sigma_{\mu\nu}\gamma_5 \Lambda^*_{S_{01}} \biggr] F^{\mu\nu}, \nn \\
\CL_{KN \Lambda^*_{S_{01}}}^{PV}&=&\frac{f_{KN \Lambda^*_{S_{01}}}}{m_k}
    (\bar{N} \gamma_\mu  \Lambda^*_{S_{01}} \partial^\mu K
       +\partial^\mu K^\dagger \bar{\Lambda}^*_{S_{11}} \gamma_\mu  N  ),
\end{eqnarray}
and for $\Sigma^*(L_{I2J}=S_{11})$ hyperon resonances
\begin{eqnarray}
\CL_{\gamma \Sigma^{0*}_{S_{11}} \Lambda} & =&-e \frac{\kappa(\Sigma^{0*}_{S_{11}} N)}{4M}
    \biggl[\bar{\Sigma}^{0*}_{S_{11}} \sigma_{\mu\nu}\gamma_5  \Lambda
+\bar{\Lambda}\sigma_{\mu\nu}\gamma_5 \Sigma^{0*}_{S_{11}} \biggr] F^{\mu\nu}, \nn \\
\CL_{KN \Sigma^*_{S_{11}}}^{PV}&=&\frac{f_{KN \Sigma^*_{S_{11}}}}{m_k}
    (\bar{N} \gamma_\mu  \vec{\Sigma}^*_{S_{11}}\cdot\vec{\tau} \partial^\mu K
   +\partial^\mu K^\dagger \bar{\vec{\Sigma}}^*_{S_{11}}\cdot\vec{\tau} \gamma_\mu N).
\end{eqnarray}
Using the above Lagrangians, we obtain
\be
\CM^{PV}= \CM^{PS} + e \frac{G_{Y^*}}{2 M (M_{Y^*}-M)} \frac{u-M_{Y^*}^2}{u-M_{Y^*}^2+i \Gamma M_{Y^*}}
\bar{u}_\Lambda \frac{1}{2} [k\!\!\!/ ,\epsilon\!\!\!/] \gamma_5 u \nn \\
\ee
with
\be
\CM^{PS}=e \frac{G_{Y^*}}{2 M }
\bar{u}_\Lambda \gamma_5 \frac{1}{2} [k\!\!\!/ ,\epsilon\!\!\!/]
\frac{p\!\!\!/ -q\!\!\!/ +M_{Y^*}}{u-M_{Y^*}^2+i \Gamma M_{Y^*}}u,
\ee
and $Y^*=\Lambda^*(S_{01}),\Sigma^* (S_{11})$.
In the above equation, the coupling constant $
G_{Y^*}= \kappa(Y^*\Lambda)*g_{KNY^*}, ~
g_{KNY^*} = \frac{M-M_{Y^*}}{m_K} f_{KNY^*}
$.

The interaction Lagrangians of $\Lambda^*(L_{I 2J}=P_{01})$ isobar  are given by
\begin{eqnarray}
\CL_{\gamma \Lambda^*_{P_{01}} \Lambda} & =&-e \frac{\kappa(\Lambda^*_{P_{01}} N)}{4M}
    \biggl[\bar{\Lambda}^*_{P_{01}} \sigma_{\mu\nu}  \Lambda
+\bar{\Lambda}\sigma_{\mu\nu} \Lambda^*_{P_{01}} \biggr] F^{\mu\nu}, \nn \\
\CL_{KN \Lambda^*_{P_{01}}}^{PV}&=&\frac{f_{KN \Lambda^*_{P_{01}}}}{m_k}
   ( \bar{N} \gamma_\mu\gamma_5  \Lambda^*_{P_{01}} \partial^\mu K
       +\partial^\mu K^\dagger  \bar{\Lambda}^*_{P_{11}}
         \gamma_\mu\gamma_5  N ),
\end{eqnarray}
and for $\Sigma^*(L_{I2J}=P_{11})$
\begin{eqnarray}
\CL_{KN \Sigma^*_{P_{11}}}^{PV}&=&\frac{f_{KN \Sigma^*_{P_{11}}}}{m_k}
   ( \bar{N} \gamma_\mu \gamma_5 \vec{\Sigma}^*_{P_{11}}\cdot\vec{\tau} \partial^\mu K
      + \partial^\mu K^\dagger
       \bar{\vec{\Sigma}}^*_{P_{11}}\cdot\vec{\tau}
         \gamma_\mu \gamma_5 N ), \nn \\
\CL_{\gamma \Sigma^{0*}_{P_{11}} \Lambda} & =&-e \frac{\kappa(\Sigma^{0*}_{P_{11}} N)}{4M}
    \biggl[\bar{\Sigma}^{0*}_{P_{11}} \sigma_{\mu\nu}  \Lambda
+\bar{\Lambda}\sigma_{\mu\nu} \Sigma^{0*}_{P_{11}} \biggr] F^{\mu\nu}.
\end{eqnarray}
From the above Lagrangians, the reaction matrix $\CM$ is obtained as
\Be
\CM^{PV}&=& \CM^{PS} + e \frac{G_{Y^*}}{2 M (M+M_{Y^*})} \frac{u-M_{Y^*}^2}{u-M_{Y^*}^2+i \Gamma M_{Y^*}}
\bar{u}_\Lambda \frac{1}{2} [k\!\!\!/ ,\epsilon\!\!\!/] \gamma_5 u, \nn \\
\CM^{PS}&=&e \frac{G_{Y^*}}{2 M }
\bar{u}_\Lambda \frac{1}{2} [k\!\!\!/ ,\epsilon\!\!\!/]
\frac{(p\!\!\!/ -q\!\!\!/ +M_{Y^*})}{u-M_{Y^*}^2+i \Gamma M_{Y^*}}\gamma_5 u,
\Ee
and $Y^*=\Lambda^*(P_{01}),\Sigma^* (P_{11})$.
In the above equation, the coupling constant $
G_{Y^*}= \kappa(Y^*\Lambda)*g_{KNY^*}, ~
g_{KNY^*} = \frac{M+M_{Y^*}}{m_K} f_{KNY^*}
$.

\subsection{Spin 3/2 Isobars}
\label{sec:spin3}

A free Lagrangian for a  spin-3/2 field
is invariant under a point transformation:
\be
N^{*\mu} \longrightarrow N^{*\mu} +\alpha \gamma^\mu\gamma^\nu N_\nu^*
\label{eq:point}~,
\ee
where $\alpha$ is an arbitrary parameter\cite{Pecc68}.
Since $N^{*\mu}$ field 
satisfies the Euler-Lagrange equation and other subsidiary conditions:
\be
 (p\!\!\!/ -M^*) N^{*\mu} (p) = 0,  ~\gamma_\mu N^{*\mu}(p) = 0 ,~
 p_\mu N^{*\mu}(p)= 0 ~,
\label{eqpt}
\ee
the transformation does not affect the spin-3/2 component of the $N^{*\mu}$, but
mixes the two classes of spin-1/2 contents of $N^{*\mu}$.
Therefore, if we want a pure spin-3/2 coupling Lagrangian, it is necessary
that the Lagrangian remains invariant under the point transformation.
Following Peccei\cite{Pecc68}, for a given Lagrangian
\be
\CL_{\rm int}=h \bar{N}^{*\mu} \Theta_{\mu\nu} N B^\nu,
\label{lspin3}
\ee
the invariant interaction Lagrangian is obtained by imposing a subsidiary
condition on the coupling matrix $\Theta_{\mu\nu}$,
\be
\gamma_\mu \Theta^{\mu\nu}=0.
\label{invO}
\ee
From starting the  $\Theta_{\mu\nu}=g_{\mu\nu}$,
we construct the $\Theta^{\mu\nu}$
\be
\Theta_{\mu\nu} =g_{\mu\nu}-\frac{1}{4} \gamma_\mu\gamma_\nu,
\ee
which satisfies the condition Eq. (\ref{invO}).

For $N_R(P_{13})$-nucleon resonance 
the general form of interaction Lagrangians is given by
\begin{eqnarray}
\CL_{\gamma N_R N}&=&\CL_{\gamma N_R N}^1 +\CL_{\gamma N_R N}^2, \nn\\
\CL_{\gamma N_R N}^1 &=& -i e \frac{C_1^{\rm em}}{2 M} \bar{N}_R^\mu \Theta_{\mu\nu} \gamma_\lambda  \gamma_5 N F^{\nu\lambda} + h.c., \nn \\
\CL_{\gamma N_R N}^2 &=& e \frac{C_2^{\rm em}}{4 M^2} \bar{N}_R^\mu \Theta_{\mu\nu}  \gamma_5   \partial_\lambda N F^{\nu\lambda} + h.c., \nn \\
\CL_{KN_R \Lambda} &=&\frac{f_{KN_R\Lambda}}{m_K}( \bar{N}_R^\mu \Theta_{\mu\nu} \Lambda\partial^\nu K
                 + \partial^\nu K^\dagger \bar{\Lambda} \Theta_{\nu\mu} N_R^\mu )~,
\label{la3N}
\end{eqnarray}
and for $\Lambda_R(P_{03})$-hyperon resonance
\begin{eqnarray}
\CL_{\gamma \Lambda_R \Lambda}&=&\CL_{\gamma \Lambda_R \Lambda}^1 +\CL_{\gamma \Lambda_R \Lambda}^2, \nn\\
\CL_{\gamma \Lambda_R \Lambda}^1 &=& -i e  \frac{C_1^{\rm em}}{2 M} \bar{\Lambda}_R^\mu \Theta_{\mu\nu}\gamma_\lambda \gamma_5 \Lambda   F^{\nu\lambda} + h.c, \nn \\
\CL_{\gamma \Lambda_R \Lambda}^2 &=& e  \frac{C_2^{\rm em}}{4 M^2} \bar{\Lambda}_R^\mu \Theta_{\mu\nu}   \partial_\lambda \Lambda F^{\nu\lambda} + h.c., \nn \\
\CL_{K\Lambda_R N} &=&\frac{f_{K\Lambda_R N}}{m_K} (\partial^\nu K^\dagger \bar{\Lambda}_R^\mu \Theta_{\mu\nu} \gamma_5 N
                 + \bar{N} \Theta_{\nu\mu}(w) \gamma_5\Lambda_R^\mu \partial^\nu
                 K),
\label{la2Y}
\end{eqnarray}
where $C_1^{\rm em}$ and $C_2^{\rm em}$ are electromagnetic coupling
constants and $f$ is a strong coupling constant. 
A Lagrangian for $\Sigma_R(P_{13})$ is easily obtained
by the following replacements of the Lagrangians in Eq. (\ref{la2Y})
\Be
\Lambda_R^\mu \longrightarrow \Sigma_R^{0*\mu} \ \ \mbox{ in $\CL_{\gamma \Lambda_R \Lambda}$}, \nn \\
\Lambda_R^\mu \longrightarrow \vec{\tau}\cdot \vec{\Sigma}_R^\mu \
\mbox{ in $\CL_{K\Lambda_R N}$}.
\Ee

Interaction Lagrangians of $D_{13}$ and $D_{03}$ resonances are obtained
by replacing the fields of $P_{13}$ and $P_{03}$ as
\be
N_R^\mu \rightarrow \gamma_5 N_R^\mu,~ 
\bar{N}_R^\mu \rightarrow - \bar{N}_R^\mu \gamma_5,
\ee
respectively.

\subsection{Spin 5/2 Resonances}
\label{sec:spin5}

The electromagnetic interaction Lagrangian of $F_{15}$-nucleon resonance is
given by
\begin{eqnarray}
\CL_{\gamma N_R N}&=&  \CL_{\gamma N_R N}^1 +\CL_{\gamma N_R N}^2 , \nn\\
\CL_{\gamma N_R N}^1 &=&  e \frac{C_1^{\rm em}}{(2 M)^2} \bar{N}_R^{\mu\nu} \Theta_{\mu\alpha} \gamma_\lambda  N \partial_\nu F^{\alpha\lambda} + H.C \nn ,\\
\CL_{\gamma N_R N}^2 &=& i e \frac{C_2^{\rm em}}{(2 M)^3} \bar{N}_R^{\mu\nu} \Theta_{\mu\alpha}  \partial_\lambda N \partial_\nu F^{\alpha\lambda} + H.C,
\end{eqnarray}
and the strong interaction Lagrangian by
\begin{equation}
\CL_{KN_R \Lambda} =i \frac{f_{KN_R\Lambda}}{m_K^2} \bar{N}_R^{\mu\nu}
  \gamma_5\Theta_{\mu\alpha} N \partial^\nu \partial^\alpha K
                 +i \frac{f_{KN_R\Lambda}}{m_K^2} \bar{N}
  \gamma_5\Theta_{\alpha\mu} N_R^\mu \partial^{\mu\nu}\partial^\alpha K.
\end{equation}
In the above Lagrangians we take the $\Theta_{\mu\nu}$ as the 
following simple form:
\begin{equation}
\Theta_{\mu\nu} =g_{\mu\nu}
\end{equation}
and $M$ is a nucleon mass. $f$ is a strong coupling constant  and $C_1^{\rm em}$ and $C_2^{\rm em}$
are electromagnetic coupling constants. The interaction Lagrangian of $D_{15}$-nucleon
resonance is obtained by the following replacement of field
\begin{equation}
N_R^{\mu\nu} \rightarrow \gamma_5 N_R^{\mu\nu}, ~
\bar{N}_R^{\mu\nu} \rightarrow -\bar{N}_R^{\mu\nu} \gamma_5.
\end{equation}
\section{Strong Form Factors and Gauge Invariance}
\label{secform}

The hadronic form factors are introduced in KP 
in order to consider effects of internal structures of hadrons
and to improve a divergent behavior of KP amplitudes
in an isobaric model as incident photon energy increases.
Mart {\it et al.}\cite{Mart95} showed that the model showing a good description
of the $N(\gamma,\ K^+)Y$ data, 
might give unrealistically large predictions for the
$N(\gamma,\ K^0)Y$ channels. This problem is alleviated by using
hadronic form factor\cite{Benn97}.
In this paper, we also introduce 
strong form factors and
investigate their effects on the KP.

It is well known, however, that the inclusion of form factors
at hadronic vertices gives rise to  gauge violation of the amplitude.
Electromagnetic(EM) current should be conserved because of 
a gauge symmetry of a system. The gauge invariance condition is
that {\it the replacing the photon  polarization vector $\epsilon$
with its four momentum $k$ makes the KP amplitude vanish}.
Without form factors, the KP amplitude obtained by a tree approximation 
is gauge invariant because the Lagrangian has been constructed as 
gauge invariant form.
However, when the strong form factors are introduced in strong vertices
the gauge invariance is violated\cite{Benn97,Habe98,Ohta89,Habe97}.
Certainly, the KP amplitude for  each $s$, $u$, and $t$ channel
resonances
is gauge invariant because the EM vertices are constructed to be
self gauge invariant.

In the Born terms, the $u$-channel amplitudes are gauge invariant
because its EM-vertex function is proportional to
$k\!\!\!/\epsilon\!\!\!/$.
However, a sum of $s$-, $t$-channel and Kroll Rudermann(KR) amplitudes
is given by
\Be
M_s^{PV}+M_t^{PV}+M_{KR}^{PV}&=&  f(M^2,s,m_K^2) e  g_{KN\Lambda}
\biggl[\frac{1+2M \mu_p}{s-M^2} +\frac{\mu_p}{M+M_\Lambda} \biggr] {\cal O}_1 \nn \\
& & +2 f(M^2,s,m_K^2) \frac{e g_{KN\Lambda}}{s-M^2}\mu_p {\cal O}_3+\CM_{\rm Born}^{viol}~,
\label{mstc}
\Ee
with
\Be
\CM_{\rm Born}^{viol}&=&\frac{4 e g_{K N\Lambda}}{(s-M^2) (t-m_K^2)}
    \bar{u}_\Lambda \gamma_5 u \nn\\
 & &\times ( f (M^2, M_\Lambda^2,t)  \epsilon\cdot q p\cdot k - f(M^2,s,m_K^2)q\cdot k \epsilon\cdot
p)\nn\\
& &+\frac{e g_{KN\Lambda}}{M+M_\Lambda}\bar{u}_\Lambda \gamma_5 \epsilon\!\!\!/ u
(  f(M^2,s,m_K^2)-f_{KR}).
\label{gviol}
\Ee
In the above equation the $f(M^2,s,m_K^2)$ and $f(M^2, M_\Lambda^2,t)$ 
mean, respectively, $KN\Lambda$ strong form factors for s- and t-channel.
The KR form factor $f_{KR}=1$
because all of the particles in the vertex are on their mass shells.
While the ${\cal O}_1$ and ${\cal O}_3$ in Eq. (\ref{mstc}) are self
gauge invariant, $\CM_{\rm Born}^{viol}$ is not
because $f(M^2,s,m_K^2)\neq\ f(M^2, M_\Lambda^2,t)$. Therefore the
Eq. (\ref{mstc}) is not gauge invariant.

Ohta contrived a prescription to  restore  the gauge  invariance\cite{Ohta89}.
He divided it into two parts
\be
\CM^{Ohta}=\CM^B +\Delta\!\CM^{Ohta}.
\label{cm}
\ee
$\CM^B$ is a sum  of three  terms which contain  the isolated nucleon(hyperon)
or pion pole and
$\Delta\!\CM^{Ohta}$ is a contact term which 
is obtained by minimal replacement
of the following most general $KN\Lambda$ vertex function\cite{Kaze59}
\Be
\Gamma(p_2,\ p_1,\  p_K)&=&\gamma_5 f_1(p_2^2,p_1^2,p_K^2)+
      \gamma_5  (p_1\!\!\!\!\!/ -M_1)  f_2(p_2^2,p_1^2,p_K^2)
\nn \\
   & & + (p_2\!\!\!\!\!/ -M_2) \gamma_5 f_3 (p_2^2,p_1^2,p_K^2) \nn \\
   & & + (p_2\!\!\!\!\!/ -M_2)\gamma_5 (p_1\!\!\!\!\!/ -M_1) f_4(p_2^2,p_1^2,p_K^2),
\label{gam}
\Ee
where $p_1,\ p_2$ and $p_K$ are momenta of the initial and final baryons 
and kaon.
Conservation of momentum in strong vertex gives a relation, 
$p_1=p_2+p_K$.
He proved the gauge invariance of the resulting $\CM^{Ohta}$
by   showing  that it  satisfies  the  Ward-Takahashi identity\cite{Ward50}.

Following Ohta's prescription, the gauge invariant Born amplitudes with PV coupling
are obtained by the following way.
Since the  $\CM^B$ in Eq. (\ref{cm}) is not gauge invariant, we can divide it
into a gauge invariant part ($\CM^{inv}$) and
a gauge violating part ($\CM^{viol}$):
\Be
\CM^B &=& \CM^{inv}+\CM^{viol},
\label{divcm}
\Ee
where $\CM^{inv}$ and $\CM^{viol}$ are given by
\Be
\CM^{inv} &=&f(M^2,s,m_K^2) e  g_{KN\Lambda}
\biggl[\frac{1+2M \mu_p}{s-M^2} +\frac{\mu_p}{M+M_\Lambda} \biggr] {\cal O}_1 \nn \\
& & + 2 f(M^2,s,m_K^2) \frac{e g_{KN\Lambda}}{s-M^2}\mu_p {\cal O}_3, \\
\label{eqgin}
\CM^{viol}&=&\frac{e g_{KN\Lambda}}{M+M_\Lambda}\bar{u}_\Lambda \gamma_5 \epsilon\!\!\!/ u
  f(M^2,s,m_K^2)+\frac{4e g_{K N\Lambda}}{(s-M^2) (t-m_K^2)} \bar{u}_\Lambda \gamma_5 u  \nn\\
& &\times( f (M^2, M_\Lambda^2,t)  \epsilon\cdot q p\cdot k - f(M^2,s,m_K^2)q\cdot k \epsilon\cdot
p)
\label{gviol2}
\Ee
, respectively.
In order to obtain the contact term ($\Delta \CM^{Ohta}$),
we generate the PV coupling vertex from Eq. (\ref{gam}) by
the following replacement
\Be
f_1(p_2^2,p_1^2,p_K^2)&\rightarrow&f(p_2^2,p_1^2,p_K^2),~ 
f_2(p_2^2,p_1^2,p_K^2)\rightarrow\frac{f(p_2^2,p_1^2,p_K^2)}{M_1+M_2}, \nn \\
f_3(p_2^2,p_1^2,p_K^2)&\rightarrow&\frac{f(p_2^2,p_1^2,p_K^2)}{M_1+M_2},~ 
f_4(p_2^2,p_1^2,p_K^2)\rightarrow0.
\Ee
Then, by  Ohta's procedure, one can easily obtain the contact amplitude:
\be
\Delta\!\CM^{Ohta}=-\CM^{viol} + f(p'^2,p^2,q^2)
 \frac{4e g_{K N\Lambda}}{(s-M^2) (t-m_K^2)} \bar{u}_\Lambda \gamma_5 u
(\epsilon\cdot q p\cdot k - q\cdot k \epsilon\cdot p).
\label{dcm}
\ee
Substituting the Eqs. (\ref{gviol2}) and (\ref{dcm}) into Eq. (\ref{divcm}) gives
the gauge invariant amplitude:
\be
\CM^{viol}+\Delta\!\CM^{Ohta}=e g_{K N\Lambda}  \bar{u}_\Lambda \gamma_5 u \frac{4}{(s-M^2)
(t-m_K^2)}
(\epsilon\cdot q p\cdot k - q\cdot k \epsilon\cdot p),
\label{restg}
\ee
where we use
\be
f(p'^2,p^2,q^2)=f(M_\Lambda^2,M^2,m_K^2)=1
\label{f1}
\ee
because the three particles are on their mass shells.
The resulting $\CM^{Ohta}$ is
\Be
\CM^{Ohta} &=& f(M^2,s,m_K^2) e  g_{KN\Lambda}
\biggl[\frac{1+2M \mu_p}{s-M^2} +\frac{\mu_p}{M+M_\Lambda} \biggr] {\cal O}_1 \nn \\
& & + 2 f(M^2,s,m_K^2) \frac{e g_{KN\Lambda}}{s-M^2}\mu_p {\cal O}_3 \nn \\
&+ &e g_{K N\Lambda}  \bar{u}_\Lambda \gamma_5 u \frac{4}{(s-M^2)
(t-m_K^2)} (\epsilon\cdot q p\cdot k - q\cdot k \epsilon\cdot p).
\label{cmOhta}
\Ee
Comparing the above equation Eq. (\ref{cmOhta}) with Eq. (\ref{mstc}),
one can see that Ohta restores the gauge invariance by neglecting the form
factor effects in the gauge violating term, which was  pointed
out by Workman {\it et al.}\cite{Work92}.

Although the Ohta's prescription is widely used to cure the gauge violation,
it has some flaws\cite{Feus98,Benn97,Habe97,Wang96}.
Firstly, the Eq. (\ref{f1}) and $p_1=p_2+p_K$ give
$p=p'+q$,
which is incompatible with momentum conservation, $p+k=p'+q$.
The normalized  form factor in  Eq. (\ref{f1}) is in unphysical region.
 Furthermore the Lagrangian derived by Ohta
is not hermitian\cite{Wang96}.

Another recipe is the prescription of
Haberzettl\cite{Habe97}.
Ohta  obtains the contact current, $\Delta\!\CM^{Ohta}$,  by treating
the three momenta $p',\ p$, and $ q$ in the strong vertex(Eq. (\ref{gam}))
as the independent variables before the photon minimally couples to the vertex.
It invokes the unphysical condition Eq. (\ref{f1}).
Haberzettl removes one variable-dependency by using the condition
$p_1=p_2+p_K$, and the 
minimal substitution is done for the other independent variables.
In this way this he showed that the total production amplitude is gauge
invariant if the bare contact currents satisfy a continuity
equation.
%
%
Following the Refs. \cite{Feus98,Habe98,Habe97}, for PV coupling,
we obtain the gauge invariant amplitude $\CM^{Habe}$:
\Be
\CM^{Habe} &=& f(M^2,s,m_K^2) e  g_{KN\Lambda}
\biggl[\frac{1+2M \mu_p}{s-M^2} +\frac{\mu_p}{M+M_\Lambda} \biggr] {\cal O}_1 \nn \\
& & + 2 f(M^2,s,m_K^2) \frac{e g_{KN\Lambda}}{s-M^2}\mu_p {\cal O}_3 \nn \\
& & + (a_1 f(M_\Lambda^2,s,m_K^2)+ (1-a_1)f(M_\Lambda^2,M^2,t)) e g_{K N\Lambda}  \bar{u}_\Lambda \gamma_5 u\nn \\
& &\times \frac{4}{(s-M^2)
(t-m_K^2)} (\epsilon\cdot q p\cdot k - q\cdot k \epsilon\cdot p).
\label{cmhabe}
\Ee
Although  $a_1$ is a free parameter,
we choose $a_1=1/2$\cite{Feus98} to reduce the number of parameters.
Unlike  the Ohta's prescription (see Eq. (\ref{cmOhta})),
the last term in Eq. (\ref{cmhabe})
is multiplied by the form factor
\be
(a_1 f(M_\Lambda^2,s,m_K^2)+ (1-a_1)f(M_\Lambda^2,M^2,t)).
\label{habeform}
\ee
As a summary of the difference between the two gauge
prescriptions, for the Ohta's prescription, the form factor effects  disappear in the
$A_2$ amplitude of Born term and for Haberzettl's one, $A_2$ is multiplied by the form
factor in Eq. (\ref{habeform}) (see Eqs. (\ref{mfi}), (\ref{cmOhta}), and
(\ref{cmhabe})).

To investigate dependence of the gauge prescription, we will perform
$\chi^2$-fitting using both Ohta's prescription and  Haberzettl's one.

We use the following  phenomenological form factors\cite{Zhan95}
\Be
f(M_\Lambda^2,s,m_K^2) &=& \frac{\Lambda^4}{\Lambda^4 +(s-M^{*2})^2}, \nn \\
f(u,M^2,m_K^2) &=& \frac{\Lambda^4}{\Lambda^4 +(u-M^{*2})^2}, \nn \\
f(M_\Lambda^2,M^2,t) &=& \left(\frac{\Lambda^2-M^{*2}}{\Lambda^2-t}\right)^2,
\label{eqn:form}
\Ee
where $\Lambda$ and $M^*$ are a free cutoff parameter and mass of the particle
on the off mass shell leg in the strong vertex, respectively.
Taking the different values $\Lambda$'s for particles with different
spins, we determine them by a $\chi^2$-fitting procedure.
The resulting $\Lambda$'s are presented in the Table
\ref{tab:result2}.

\section{Fitting Strategy and Determination of Parameters}
\label{secfit}

\subsection{General Scheme}
In this section, we present our method to select particles
included in our calculation and to determine the coupling
constants by a fitting procedure.
The  fitting procedure was done by taking into account
all possible contributions
of  26 intermediate particles listed in Table \ref{nrlist1}.
Except the particles contributing to the Born terms, we have  22 resonance
particles: two spin-1 mesons,
four(seven) spin-$\frac{1}{2}$ nucleon(hyperon) resonances,
three(three) spin-$\frac{3}{2}$ nucleon(hyperon) resonances, and
two spin-$\frac{5}{2}$ nucleon resonances.
A spin-$1/2$ particle
has a coupling constant for each strong and electromagnetic vertex,
but a particle with spin $J >1/2$ has one strong  and
two electromagnetic coupling constants.
In the corresponding amplitude, the strong
and electromagnetic coupling constants appear as a multiplied form.
Therefore,  the multiplied coupling constant is a free parameter to be
determined in the  analysis for the KP.

All parameters except the anomalous magnetic moments of the proton and
$\Lambda$, and $\Sigma-\Lambda$ transition magnetic moment, are not exactly
known, so that they will be determined by fitting the total cross section,
differential cross
section, $\Lambda$-polarization asymmetry ($P$),
target-polarization asymmetry ($T$), and radiative kaon capture ratio
to their experimental data. However,
because the particles strongly interfere with each other,
it is not easy,  within the
physically acceptable ranges, to determine them
from the present available data. Particularly, we can not set up 
physical boundaries for them because we have no experimental information
about the electromagnetic coupling constants of hyperon
resonances.
Moreover, if we perform the fitting procedure without constraint conditions for the coupling constants,
the minimum occurs at the unphysically large values for the hyperon resonances.
To solve the problem we consider a nonrelativistic quark
model(NRQM) prediction for the coupling constants of hyperon
resonances.

Adelseck  {\it et al.}'s analysis of the existing differential cross
section data for the KP up to $E_\gamma^{\rm Lab}=$1.4 GeV yield some simple
models, which was consistent with the broken $SU(3)$
 symmetry,  by the excluding the Orsay's 22 data\cite{Deca70} which  reveal the internal
 inconsistency\cite{Adel90}.
On the same lines, we use the same data set as  used in Ref.
\cite{Adel90} and the data in Ref. \cite{Bock94}, i.e., we use 
251 the experimental
data: 197  differential cross
section, 21 total cross section, 30 $\Lambda$-polarization asymmetry,
3 target-polarization asymmetry, and 1 radiative kaon capture ratio data.
Our fitting procedure is fulfilled by
using a CERN library package MINUIT\cite{Minu81}.

For the simple model of the KP,
we investigate a sensitivity of a $\chi^2$ value for
coupling constants of particles and discard the particle played minor roles
in the $\chi^2$ sensitivity, for example, ($N5$ and $N9$).
We start the fitting by including all the particles 
listed in Table \ref{nrlist1},
whose coupling constants are
constrained by the following way.

\subsection{Leading Coupling Constants}
In the Born terms, electromagnetic coupling 
constants such as the anomalous magnetic
moments of a proton and  $\Lambda$, and
$\Sigma-\Lambda$ transition magnetic moment 
have been accurately given by\cite{PDG94,PDG98}, 
using de Swart convention\cite{Swar63},
\be
\ba{ccc}
\kappa_p= 1.79,&\ \ \ \ \kappa_\Lambda=-0.6138,& \ \ \ \
\kappa(\Sigma\Lambda)=1.61.
\ea
\ee
The strong coupling constants $g_{KN\Lambda}$ and $g_{KN\Sigma}$
have not been well determined. The broken $SU(3)$ symmetry predicts
\be
g_{KN\Lambda} = -\frac{1}{\sqrt{3}} (3-2 \alpha_D) g_{\pi NN}, 
~~g_{KN\Sigma}  = (2 \alpha_D-1) g_{\pi NN},
\ee
where $\alpha_D$ is a fraction of $D$-type coupling.
Using the experimental value\cite{Adel90}:
\be
\frac{g_{\pi NN}^2}{4 \pi}=14.3\pm 0.2, \ \ \ \
\alpha_D =0.644 \pm 0.006 ~,
\ee
we obtain $SU(3)$ predicting coupling constants:
\be
\frac{g_{KN\Lambda}}{\sqrt{4 \pi}}= -3.74, 
~~\frac{g_{KN\Sigma}}{\sqrt{4 \pi}} =  1.09.
\label{eqgfix}
\ee
Considering about $20 \%$ breaking of $SU(3)$ symmetry, however, 
gives ranges of the leading coupling constant
\be
-4.3 \leq \frac{g_{KN\Lambda}}{\sqrt{4 \pi}} \leq -3.0,~~ 
0.9 \leq \frac{g_{KN\Sigma}}{\sqrt{4 \pi}} \leq  1.3 ~.
\label{eqgfree}
\ee
Although some works\cite{Adel90,Davi96} are consistent with Eq.
(\ref{eqgfree}), other many papers are inconsistent with the
values in Eq. (\ref{eqgfree}). 
J. Cohen \cite{Cohe89}pointed out that KP is not
appropriate  to extract the leading coupling constants from it unambiguously.
However, as pointed out in our previous paper \cite{Cheo96}, 
to determine them we have to go down to lower
energy region i.e. near threshold, where most of the resonances
are expected to play minor roles in this reaction. Near coming
data at this range would clarify this problem.

In the energy
region considered here, on the contrary, we had better pay
attention to the coupling constants of the resonance 
by fixing the leading coupling constants.
Of course, we check the dependence of our final result on the
values of the leading coupling constant in the following way.
We resort to the  fitting procedure by two following methods:
i) we fix them as in Eq. (\ref{eqgfix})
 and ii) vary them
in the range in Eq. (\ref{eqgfree}). For the two cases, we obtain
nearly same $\chi^2$ as shown in Table\ref{tab:result1}.
Therefore, in our calculation, we fix the coupling constants,
$g_{KN\Lambda}$ and $g_{KN\Sigma}$, to the values in Eq. (\ref{eqgfix}).

\subsection{Spin 1 Mesons}

For the spin-1 mesons, the coupling constants and their 
broken $SU(3)$ symmetry predicting ranges are given
in our previous paper \cite{Cheo96}. Although
we took the sign of $G_{K^* K^+ \gamma}$ as plus, in fact, the sign is not
determined.
In addition, the t-channel resonances are related to the s- and u-channel
of high spin resonances by duality, so that we can not strictly apply 
the $SU(3)$ restriction
to the coupling constants of spin-1 meson. For that reason, we
allow them to vary freely.

\subsection{Nucleon Resonances}
As for nucleon resonances, we take the following steps.
If the experimental branching ratios 
for both strong and electromagnetic decay widths  are given 
(see Table \ref{nrlist2}),
we limit their coupling constants to the range extracted by the branching
ratios. But, as in Table \ref{nrlist2}, 
if only one is available, we can not reduce
the number of free parameters.
Some of the experimental strong decay widths for the resonance 
show a range (second column in Table \ref{nrlist2}). For the case, we
choose an intermediate  value on the given range. And then, using 
the equation for
strong decay widths given in ref.\cite{hanphd}, we obtain the strong
coupling constants  which are shown in the 5th column in Table
\ref{nrlist2}.
Likewise to the strong decay widths, electromagnetic coupling constants can be
deduced as shown in Appendix.

Among the nucleon resonances with $J=\frac{3}{2}$,
we exclude the $N5(1520) D_{13}$ because at the $\chi^2$ minimum 
its coupling constants
are unphysically large. Moreover it negligibly affects the
$\chi^2$. According to the previous works\cite{Davi96,Sagh97},
the spin-$\frac{5}{2}$ particle is very important to reproduce the nodal
structures of the spin observables.
In our calculation, we consider three cases:
i) including both $N9(1680) \frac{5}{2}^+$ and $N8(1675) \frac{5}{2}^-$,
ii) including only $N8(1675) \frac{5}{2}^-$, and iii) including
only $N9(1680) \frac{5}{2}^+$.
For i) and ii) cases, we obtain almost same results, but
for ii) and iii) cases, the former gives better result than the latter.
Therefore,
for a more simpler model space we adopt the case ii).

\subsection{Hyperon Resonances}
In case of hyperon resonances, the experimental electromagnetic decay
widths are not given except for the
$\Lambda 6(1520)$, while the strong coupling 
constants of the hyperons can be determined from
the experimental
strong decay ratios\cite{PDG94,PDG98}(see Table \ref{nrlist2}). 

The electromagnetic coupling constants for the hyperon
resonances given in column E of Table 
\ref{tab:result1} are calculated from the  resonance
coupling amplitudes given in Appendix using the mixing angles for
 $\Lambda 3$, $\Lambda 4$,
$\Sigma 2$, $\Lambda 6$ and  $\Lambda 7$ coming 
from the NRQM \cite{Isgu77,Juri83} in Table \ref{tab:nrqA1} and
their resonance coupling amplitudes 
$A_{3/2}$ and $A_{1/2}$ in Table \ref{tab:nrqA2}.
The resultant resonance
coupling amplitudes are given in Table \ref{tab:nrqA3}.

After the procedure, we do a fitting procedure
without
any limitations about their coupling constants.
In this case, as mentioned above
the minimum of $\chi^2$ takes place at unphysically large coupling constants,
so that we need conditions confining them.

We adopt the following two methods:
For the first case, we extract the strong coupling constants  from experimental
strong decay widths,
and calculate the electromagnetic 
coupling constants from the NRQM-predicting electromagnetic
decay widths. The resulting coupling constants are listed in the
column E in Table \ref{tab:result1}. In this case, we also allow 
the possibility for the
coupling constants to have  opposite signs.
The 2nd case is that we use the strong coupling constants 
of the 1st case, but 
allow variations of the electromagnetic coupling constants in the range where
their electromagnetic decay ratios are assumed to take 
values less than or equal to
$1\%$\cite{PDG94,PDG98}.

For the 2nd case, we don't find the stable minimum points in the range,
i.e., the minimum of $\chi^2$ is located at boundary values of the coupling
constants. However, as shown in the Table \ref{tab:result1}, we obtain the almost the same $\chi^2$
for the two cases.
By the reason, we fix the coupling constants of negative parity hyperon resonances
to those obtained in the first case.
For $ \Lambda 6$, $\Lambda 7$, and $\Sigma 3$, we
can see that including them
gives little improvement of the $\chi^2$ (compare the column E with G
in the Table \ref{tab:result1}). For a simpler model space, therefore,
we exclude them in our calculation.

For even parity hyperon resonances,
since we do not have any information even from the NRQM calculations about
the electromagnetic  decay widths, we perform the fitting procedure
 according to the above 2nd case.
Likewise to the above case, the minimum occurs at the boundary values of the
parameters, so that we can not determine their coupling constants. However,
we obtain the almost same values of $\chi^2$: $\chi^2/N=0.93$ for the cases 
with and
$\chi^2/N=0.97$ without  these particles where $N$ is the number of data. 
From the reason,
these particles are also 
excluded in our calculation for a simpler model space.

Consequently,  the resonance particles included in our calculation are
$K^*$, $K_1$,  and 7 nucleon resonances ($N1$, $N2$, $N3$, $N4$, $N6$, $N7$,
$N8$),and 4 hyperon resonances ($\Lambda1$, $\Lambda3$, $\Lambda4$,
$\Sigma2$).
Using them we do a fitting procedure for the following 6 cases.
The first four cases are to exploit 
PV-coupling scheme(PS-coupling scheme) with Haberzettl's and Ohta's
prescriptions to restore the gauge violation, denoted as 
Habe(PV), Habe(PS), Ohta(PV), and Ohta(PS), respectively.
The Noform(PV) and Noform(PS) are the cases using PV- and PS-coupling schemes
without form factors.
The obtained parameters for the above 6 cases are listed in Table
\ref{tab:result1}.

\section{Results and Discussions}
\label{sec:result}

Our results are analyzed for the above 6 cases explained in Section
\ref{secfit}. Resultant coupling constants,
cutoff parameters in form factors, and $\chi^2/N$ are tabulated in
Table \ref{tab:result2}. The $\chi^2$ per particle
for each physical observables is also presented in Table \ref{tab:result3}.

Table \ref{tab:result2} and \ref{tab:result3} show that
all cases, except for the Noform(PV), 
give a good $\chi^2$, which means that
 our model space is quite reasonable to explain this reaction.
In specific, Table \ref{tab:result3} shows that introducing the form
factors yields more improved results than the case without them.
Although the form factors seem to improve the total
$\chi^2$, the improvement is significant only in the total and differential
cross sections, but not in the polarization observables.
In the  target polarization asymmetry, Noform(PS) is better than Habe's and 
Ohta's.

Another important result is that a larger difference between the PS-
and PV-coupling schemes appears in the target polarization
asymmetry (T)(see Table \ref{tab:result3}). Unfortunately, 
we cannot decisively tell which coupling scheme is the better, 
because there exist only 3 data points available.

\begin{figure}
\begin{center}
\leavevmode
\epsfxsize=5.1in
\epsfysize=4.5in
\epsfbox{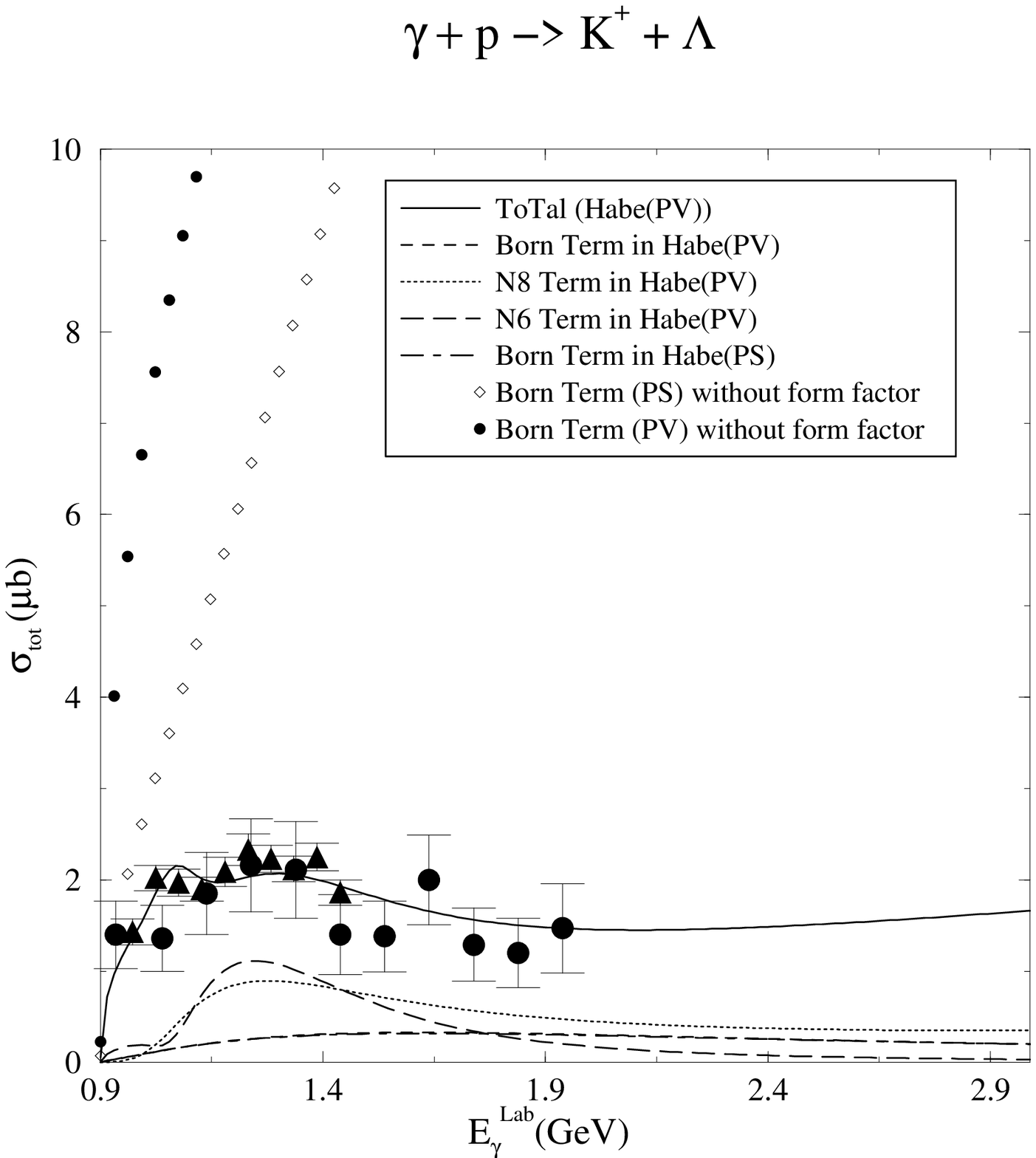}
\end{center}
\caption[The total cross
sections for dominant contributions (N6 an N8) and for PV- and PS-Born
terms with, and without form factors.]
{ The total cross
sections for dominant contributions (N6 and N8)and for PV- and PS-Born
terms with and without form factors.}.
\label{fig:partial-tot}
\end{figure}

The other interesting result is concerned to the
relation between the hadronic form factor and the difference
between the PS- and PV-coupling schemes.
When form factors are not considered in both cases,
PS scheme is superior to PV one, 
but attaching the form factor does not give any discernible difference
between the two coupling schemes (see Table \ref{tab:result3}),
which can be explained by the following facts:
{\it The difference between the two coupling schemes is shown up only in the
 spin-1/2 particles' contributions} (see Section \ref{Clag}).
Since the hadronic form factor in Eq. (\ref{eqn:form}) descends rapidly as
the momentum square of the exchanged particle goes away from its mass shell, 
effects of the form factor on the particle far away from its mass shell
is much larger than those on near mass
shell, where the
difference remains with a little diminution.
Among the spin $\frac{1}{2}$ particles, 
although the N2,  N3, and N4 have their poles
near threshold or in the energy region considered in our study,
i.e., far away from its mass shell, 
their coupling constants are so small that the difference is negligible. 
Another main contribution of spin-1/2 particles is the nucleon pole in 
the Born
terms which are near on mass shell. But, as mentioned above, the
form factor makes such a difference disappear due to the near mass shell.
To clear up this point, we depict in Fig. \ref{fig:partial-tot} total cross
sections for PV- and PS-Born
terms with, and without form factors.
While the filled circle(Noform(PV)) and unfilled diamond(Noform(PS))
show wide difference,
there is no difference between dashed curve (Habe(PV)) and 
dot-dashed curve(Habe(PS)).
If one reminds that it is inescapable to take the form factors
into account as far as we are based on the effective Lagrangian
under the tree approximation, one has to try to find
some other physical quantities to understand the difference between
the two coupling schemes.

Since a kaonic hydrogen atom has a mass of 1432 MeV, it is
expected that the $\Lambda 1$ strongly influences on 
the radiative kaon capture.
Therefore, the  branching ratio for the radiative capture of a stopped
kaon, $R_{\gamma \Lambda}$, has been studied in terms of the role
and nature of the $\Lambda 1$ \cite{Work88,Burk85,Jong92}, independently 
of the kaon
photoproduction. However, as mentioned in previous Section
\ref{seckcp}, the radiative kaon capture is related with the kaon
photoproduction through a crossing symmetry. Any realistic model 
has to explain simultaneously both reactions.
In the results of $R_{\gamma \Lambda}$ displayed in Table
\ref{tab:result4}, only Habe(PV) and Habe(PS)  fairly agree with the 
experiment.
In the Habe(PV) the $\Lambda 1$ contribution is about 24\% of the
total ratio. To say clearly about the nature of the
$\Lambda 1$, one needs more precise information on the contributions from
other particles.

\begin{figure}
\begin{center}
\leavevmode
\epsfxsize=5.2in
\epsfysize=6.0in
\epsfbox{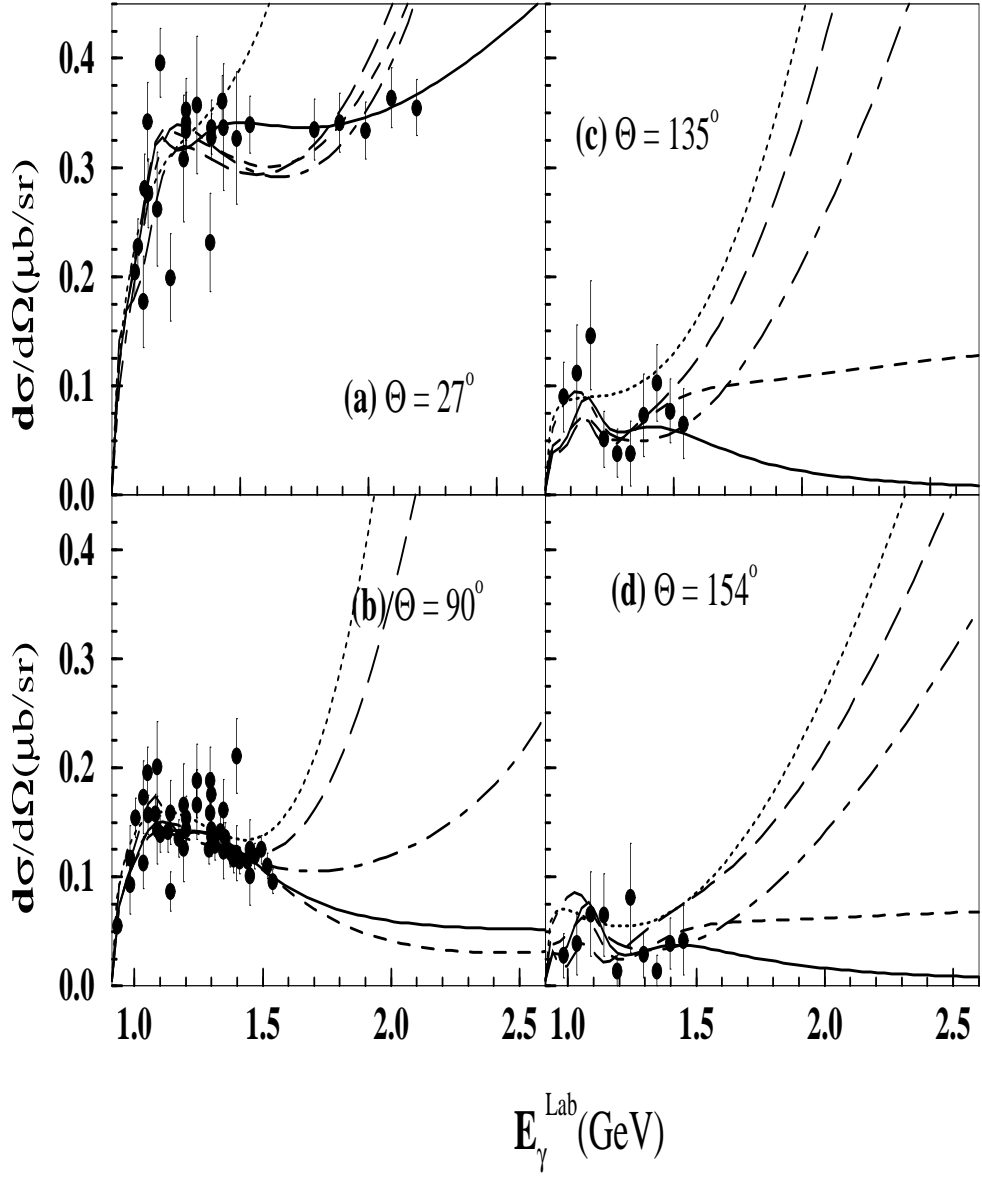}
\end{center}
\caption[The differential cross sections for fixed angle.]
{The differential cross sections for fixed kaon angles:
$\theta=$
(a) $27^\circ$, (b) $90^\circ$, (c) $135^\circ$, and (d) $154^\circ$.
Here we present the Habe(PV) with solid curve, the Ohta(PV) with dashed one.
The long dashed and dot-dashed curves  show the results of Noform(PV) and Noform(PS),
respectively.
The results of model 1 (AS1) in Ref. \cite{Adel90} are given 
by the dotted curves.
}
\label{fig:dif-e}
\end{figure}

\begin{figure}
\begin{center}
\leavevmode
\epsfxsize=5.4in
\epsfysize=6.0in
\epsfbox{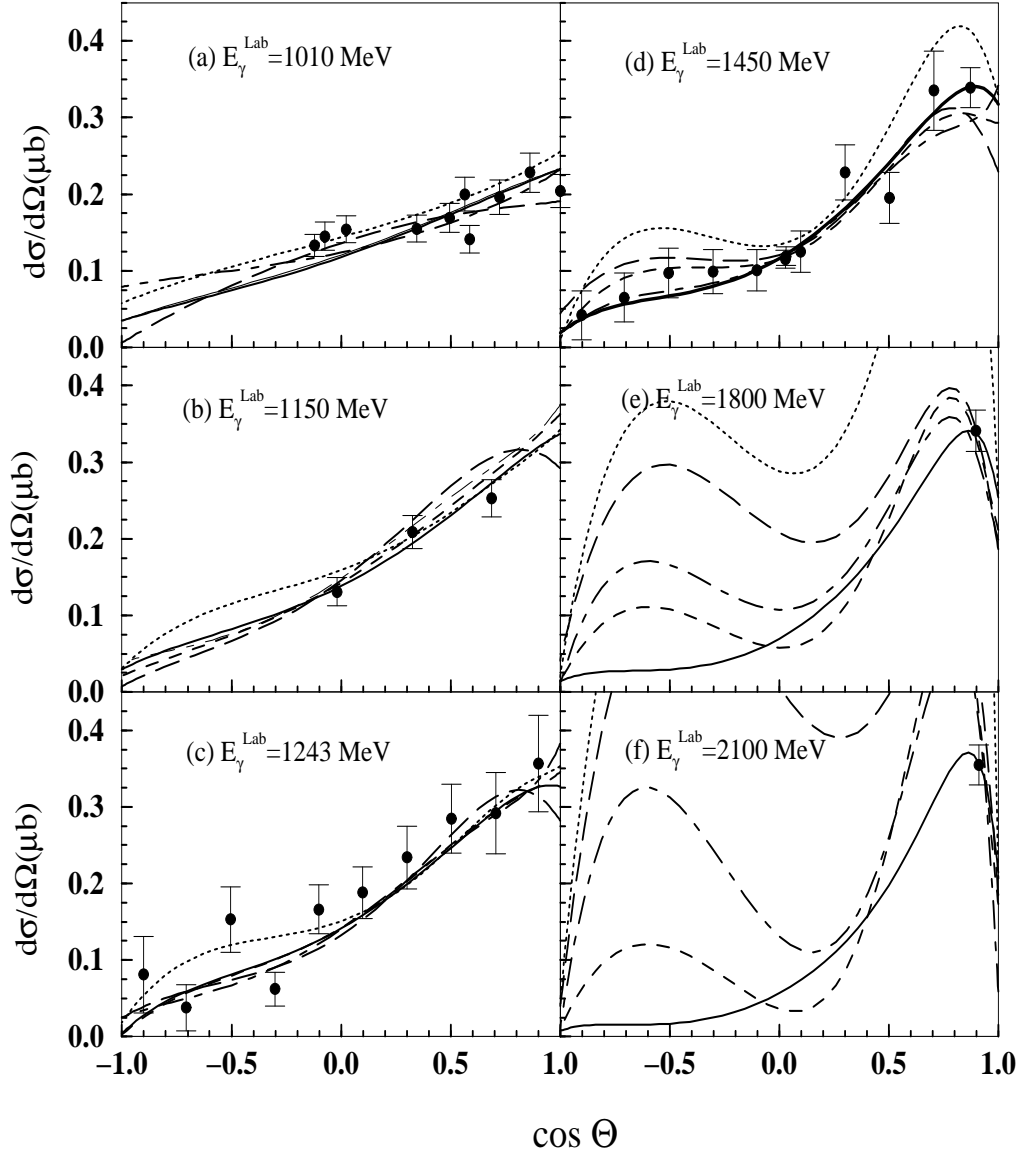}
\end{center}
\caption[The differential cross sections at fixed energy.]
{The differential cross sections at the fixed photon energy:
$E_\gamma=$
(a) $1.01$, (b) $1.15$, (c) $1.243$, (d) $1.45$, (e) $1.8$, and (f) $2.1$ GeV.
The lines are same as in Fig. \ref{fig:dif-e}. }.
\label{fig:dif-th}
\end{figure}

Fig. \ref{fig:dif-e}(a)--(d)  and Fig. \ref{fig:dif-th}(a)--(f) show 
the differential cross sections
for fixed kaon angles and for fixed incident photon energies, respectively.
As mentioned above, since the PV and PS versions in each model give almost same
results, we present only results for Habe(PV) and Ohta(PV) in the figures.

As shown in Fig. \ref{fig:dif-e} (a), at $\theta=27^\circ$,
AS1 overpredicts the data in the energy region, $E_\gamma > 1.5$ GeV
and others reproduce the experimental data up to 2.1 GeV well.
As the kaon production angle, $\theta$, increases,
they show wide differences in the high $E_{\gamma}$ range 
where no data points exist, 
although  all our models reproduce the existing experimental data.
These results reveal 
that the strong form factor, as expected, prevents the Habe(PV) and Ohta(PV)
from diverging at high photon energy $E_\gamma^{\rm Lab}$.

\begin{figure}
\begin{center}
\leavevmode
\epsfxsize=5.1in
\epsfysize=4.5in
\epsfbox{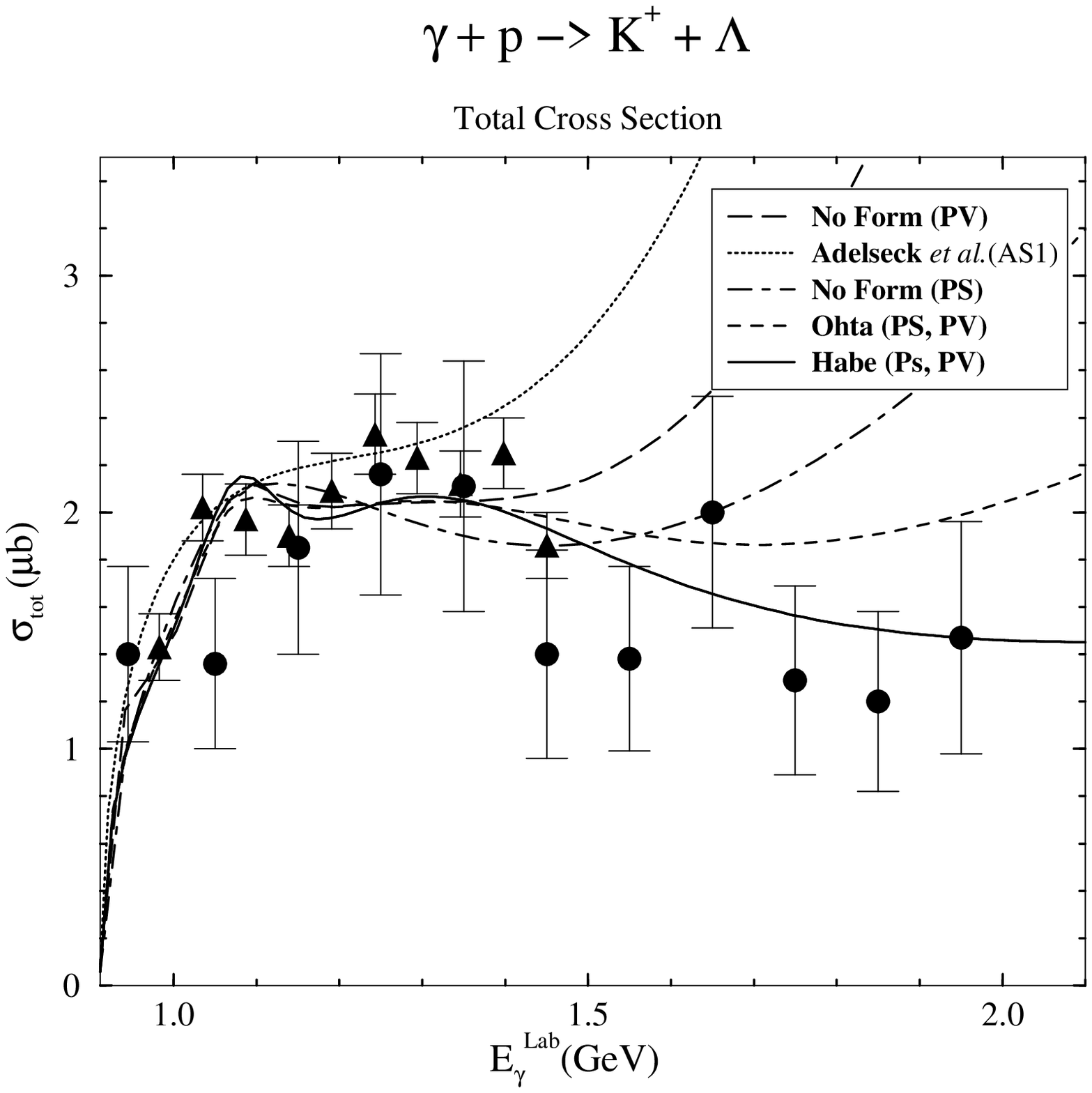}
\end{center}
\caption[The total cross sections up to energy $E_\gamma$=2.1 GeV.]
{The total cross sections up to energy $E_\gamma$=2.1 GeV; lines as in Fig. \ref{fig:dif-e}. }.
\label{fig:totcs}
\end{figure}

The  $\theta$ dependence of
the differential cross sections is depicted in Fig. \ref{fig:dif-th}.
All curves in Fig. \ref{fig:dif-th}(a)--(d)  show good predictions in
for all angles,
but the high energy results ($E_\gamma^{\rm Lab} > 1.5$), 
Fig. \ref{fig:dif-th}(e)--(f),
show significant differences for backward angles.
While the Habe(PV) monotonously decrease as the $\theta$ increases,
the others show higher peaks at backward angles, which  make
their total cross sections in Fig. \ref{fig:totcs} overpredict at high energy.
Forthcoming data at high $E_{\gamma}$ region would make it clear which
model is more realistic. 

The total cross sections are shown in Fig \ref{fig:totcs}.
The experimental  data are almost reproduced well by Habe(PV) up to 2.1
GeV.
The AS1 and Noform(PV) overpredict above 1.5 GeV and Noform(PS) and Ohta(PV)
show larger values than experimental data  in the energy higher than 1.7 GeV.
From the above analysis of the total cross section, we can conclude that the
finite size of internal structure of hadron should be considered in
the isobaric model based on effective Lagrangians as far as we stick on the
effective Lagrangian theory.

The analysis of angular distribution of polarization observable data and their
nodal trajectory offers a potentially powerful means for investigating the underlying
dynamics of KP. Therefore, the polarization observables are
intensively studied\cite{Sagh96,Sagh97}.

\begin{figure}
\begin{center}
\leavevmode
\epsfxsize=5.2in
\epsfysize=6.0in
\epsfbox{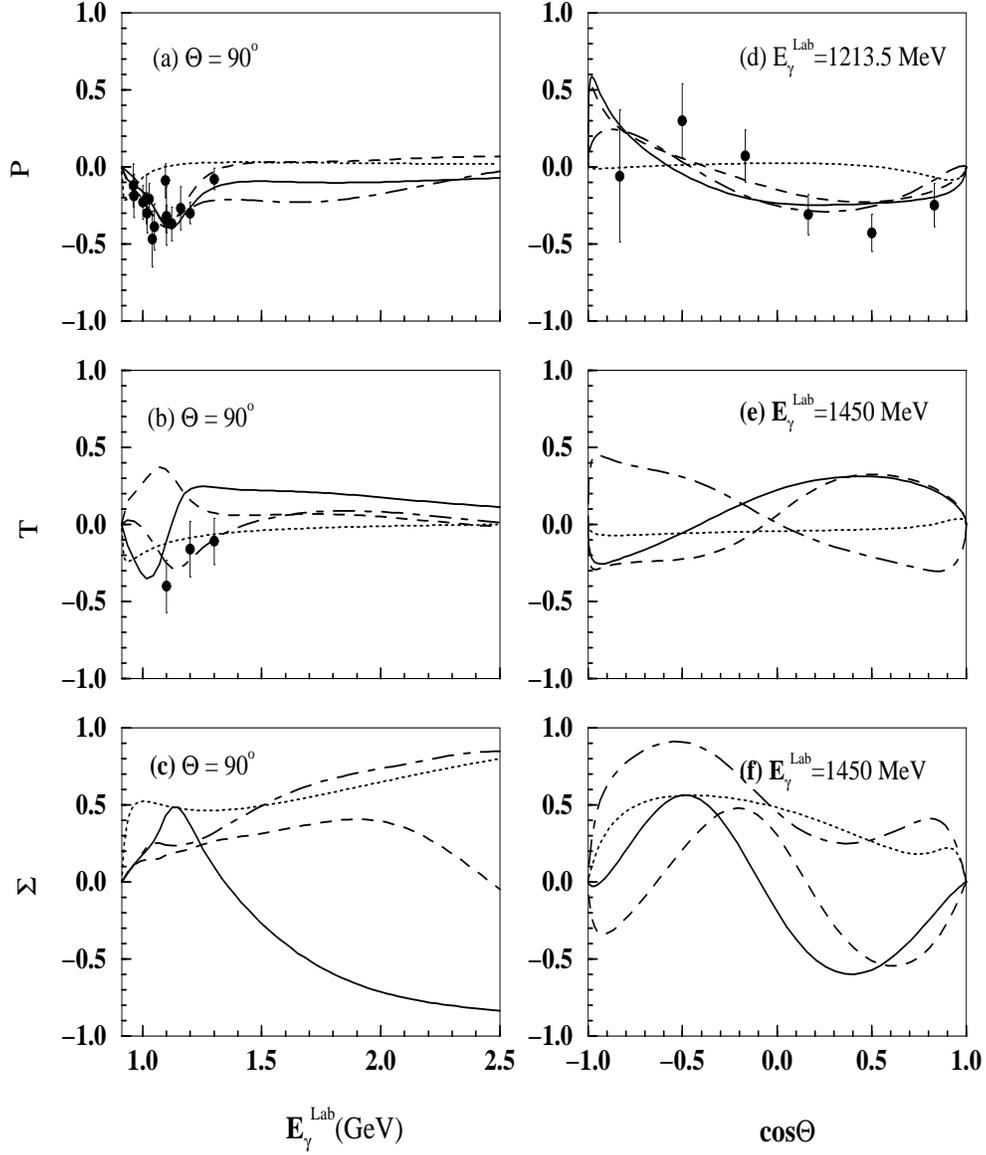}
\end{center}
\caption[The single polarization observables]
{The single polarization asymmetries: 
$\Lambda$-polarization asymmetry (P),
target-polarization asymmetry (T), and beam-polarization asymmetry ($\Sigma$);
 (a),(b), and (c) are for angle fixed at $\theta$=  $90^o$, and
$E_\gamma$= (d) 1.2135 GeV, (e) 1.450 GeV, and (f) 1.450 GeV for the photon energy fixed;
  lines as in Fig. \ref{fig:dif-e}. }
\label{fig:pol}
\end{figure}

In Fig. \ref{fig:pol}, (a) and (d), (b) and (e), and (c) and (f) represent
the results of the $\Lambda$-, target-, and beam-polarization asymmetry, respectively.
In  Fig. \ref{fig:pol}(a), we plot our fitted curves for  angle $\theta=90^\circ$.
Except for the AS1,
others show nearly same behavior and reproduce the experimental data
well.

The results for target polarization asymmetry ($T$) are displayed in  Fig. \ref{fig:pol}(b) and
Fig. \ref{fig:pol}(e)
for $\theta=90^\circ$ and $E_\gamma=1.45$ GeV, respectively.
For the target polarization asymmetry $T$, our calculations give
significantly different predictions each other(see  Fig. \ref{fig:pol}(b)).
Although the Noform(PS) and Habe(PV) give similar behaviors, the former gives slightly
better result than the latter. However, Ohta(PV) shows different sign from the data. In
 Fig. \ref{fig:pol}(e), Habe(PV)  and Ohta(PV) show nearly same behaviors, but
Noform(PS) shows opposite sign to Habe(PV) and Ohta(PV) in the whole angle.

For beam polarization asymmetry, $\Sigma$,  we present our predictions
at $\theta=90^\circ$ and $E_\gamma=1.45$ GeV in  Fig. \ref{fig:pol}(c) and
 Fig. \ref{fig:pol}(f), respectively.
Since no experimental data exists for $\Sigma$, all calculations are just the predictions
and show large difference.

Unlike the cross sections, the polarization observables show
salient
differences for each model even in low energy region.  Therefore,
we expect that forthcoming data for the polarization observables could 
single out a correct
model appropriate to this reaction.

\section{Conclusions}
\label{sec:con}

In this study, using the isobaric model, we investigate the kaon 
photoproduction,
$\gamma +p \rightarrow K^+ +\Lambda$. Our study are performed by
fitting our theoretical total cross sections, differential cross sections,
$\Lambda-$ and target-polarization symmetries
to 252 experimental data points. The SU(3)
predicting values
for $g_{KN\Lambda}$, $g_{KN\Sigma}$ are used and the coupling constants of the
resonances are varied to find the least $\chi^2$ by using the constraint
permitted by all available experimental or NRQM predicting values for
the strong and electromagnetic decay widths.
Through a systematic fitting procedure, from all the resonances
listed in Table \ref{nrlist1}, we select the following 13
resonances
\Be
\ba{llll}
     K^*(892),      &    K_1(1270),    &                   &                    \\
      N1(1440)(P_{11},)& N2(1535)(S_{11})&N3(1650)(S_{11}), &N4(1710)(P_{11}),      \\
     N6(1700)(D_{13}),     &   N7(1720)(P_{13}),  &                   &           \\
  N8(1675) (D_{15}), &                   &                   &                    \\
  \Lambda 1(1405)(S_{01}),  & \Lambda 3(1670)(S_{01}), & \Lambda 4(1800)(S_{01}), & \Sigma 2(1750)(S_{11}). \\
 \ea\nn
\Ee

Using  the resonance terms together with the Born terms,
 we intensively examine a dependence on coupling types(
  pseudovector coupling and pseudoscalar coupling)  and the effect 
of hadronic form factors.
The hadronic form factors are  introduced in a gauge invariant manner
by both
Haberzettl's and Ohta's  prescriptions.
From our analysis, the following results are obtained
\begin{enumerate}
\item When the hadronic form factors are introduced, variation
  of the leading coupling constants ($g_{KN\Lambda}$, $g_{KN\Sigma}$), 
within the broken SU(3) symmetry predicting
range, negligibly affect the $\chi^2$. We thus fix them as SU(3)
values: $g_{KN\Lambda}/\sqrt{4 \pi}=-3.74$, $g_{KN\Sigma}/\sqrt{4 \pi}=1.09$

\item Without introducing the hadronic form factors,
pseudoscalar model is superior  to pseudovector model in predictions.

\item Hadronic form factors fairly diminish the difference between
the pseudovector  and pseudoscalar coupling scheme.

\item The Haberzettl's gauge prescription gives better
results than Ohta's.

\item The hadronic form factors  improve results,  which is largely due to the
 improvement only in cross sections, while the  improvement is negligible 
in the polarization observables.

\item The analysis about target polarization data shows
  a striking difference between PV and PS
 models.

\item For a radiative
kaon capture ratio, $R_{\gamma\Lambda}$, only Haberzettl's prescription reproduces the data
well.
\end{enumerate}

Finally, for physical observables, 
the experimental data of the KP are well reproduced
in our model space and that the hadronic
form factors yield more improved results than those of the case without them. 
Although the study about hadronic form factors, which is attached to the
vertex with an off-shell
meson leg, has been done well, the information for the form factor
at the vertex with
 the off-mass shell nucleon(or hyperon) leg is scarcely known.
It is necessary to investigate the form factor through 
the more elaborate phenomenological
and  microscopic studies.
As mentioned in other 
works\cite{Davi96,Sagh96}, our analysis for the polarization observable
shows a striking model dependence in high $E_{\gamma}$ region, 
so that the forthcoming data of TJNAF, ESRF,
and ELSA, would sort out the correct one  from the existing models.
Since our model can be  applied directly to the reaction, $\gamma + p
\rightarrow K^0 +\Lambda$,  future experiments on this
reaction will also verify a validity of our model.
The kaon electroproduction is also under progress, in which
the $Q^2$ dependence of $d\sigma_L/d\sigma_T$ differs
significantly from current theoretical predictions\cite{Nicu98}.

\acknowledgements{M.K.Cheoun is grateful to Prof. Shin Nan Yang and 
Prof. S.S.Hsiao at National Taiwan University for their hospitality, 
and Dr. Biyan Saghai for his valuable comments. 
This work was supported in part by KOSEF and the
Basic Science Research Institute Program, Ministry of Education of Korea,
No BSRI-98-   .}

\section{Appendix : Radiative decay of resonances }
\label{sec:emdecay}
In this appendix we present the couplings of baryon (hyperon) resonances 
to the $\gamma N$ or $\gamma \Lambda$, which can be studied in reactions like
\be
\gamma + N \rightarrow N^* \rightarrow \pi + N, K + \Lambda,...
\ee
A partial-wave analysis of these formation processes is the
standard technique to determine the coupling constants,
$g(N^* N \gamma)$.
 The resonance coupling
amplitude for nucleon resonance
$A_\lambda^{J^P}$, in terms of the helicity amplitude of the KP, is defined by
\begin{equation}
A_\lambda^{J^P} =\frac{1}{4 \pi (2 J+1)} \, \left(\frac{1}{(2 J+1) \pi} \frac{k}{q} \frac{M}{M^*}
\frac{\Gamma_K}{\Gamma^2}\right)^{-1/2}
\, \frac{\sqrt{2 M M_\Lambda}}{W \, d_{\lambda 1/2}^J (\theta)} \,{\mbox {\rm Im}}
\biggl( -{\cal M}_{-1/2,1-\lambda}^{\lambda_\gamma =1}\biggr)~,
\end{equation}
where $\Gamma$ is a total width of the resonance, $\Gamma_K$ is 
a strong decay width
of $N^* -> \Lambda+K^+ $.
$M$ and $M^*$ are proton and  resonance masses respectively.
The helicity amplitude of kaon photoproduction 
${\cal M}_{-1/2,1-\lambda}^{\lambda_\gamma =1}$ is
defined as
\begin{equation}
{\cal M}_{-1/2,1-\lambda}^{\lambda_\gamma =1}=\epsilon^{+1}_\mu
\bar{u}_\Lambda^{-1/2} {\cal M}^\mu u_p^{1-\lambda}~,
\label{helamp}
\end{equation}
where $\lambda_\gamma$=helicity of a photon
and $\lambda$=total helicity, and $\epsilon$ is a photon polarization vector. 
All kinetic variables have values on mass shell of the resonance.
Since the resonance has
definite parity, only one(two) for $J=\frac{1}{2}$($J=\frac{3}{2}$)
is independent. Therefore, for a resonance with $J=\frac{1}{2}$$(J=\frac{3}{2})$
has independent helicity state $\lambda=\frac{1}{2}$ ($\lambda=\frac{3}{2}$ and
$\lambda=\frac{1}{2}$).

For $N^{*}(\frac{1}{2}^\pm)$ the resonance coupling amplitude is given by
\be
A_{1/2}^{\frac{1}{2}^\pm}= \pm e \kappa(N^* N) \sqrt{\frac{k}{2 M (E+M)}}.
\ee
$N^*(\frac{3}{2}^\pm)$ and $N^*(\frac{5}{2}^\pm)$ have two  resonance
coupling amplitudes.
For each resonance $N^*(P_{13})$ they are given by
\Be
A_{3/2}^{\frac{3}{2}^+} &=&\frac{e}{16 M^2}\sqrt{\frac{2 k M^*}{M}}
    (4 C_1^{\rm em} M +C_2^{\rm em} (M^*-M)), \nn \\
A_{1/2}^{\frac{3}{2}^+} &=&\frac{e}{16 M^2}\sqrt{\frac{2 k M^*}{3M}}
 \left(4 C_1^{\rm em} \frac{M^2}{M^*} -C_2^{\rm em} (M^*-M)\right),
\Ee
for $N^*(D_{13})$, and
\Be
A_{3/2}^{\frac{3}{2}^-} &=-&\frac{e}{16 M^2}\sqrt{\frac{2 k M^*}{M}}
    (4 C_1^{\rm em} M -C_2^{\rm em} (M^*+M)), \nn \\
A_{1/2}^{\frac{3}{2}^-} &=&-\frac{e}{16 M^2}\sqrt{\frac{2 k M^*}{3M}}
 \left(4 C_1^{\rm em} \frac{M^2}{M^*} -C_2^{\rm em} (M^*+M)\right),
\Ee
for $N^*(F_{15})$
\Be
A_{3/2}^{\frac{5}{2}^+} &=& 2 e f_1 \left( \frac{C_1^{\rm em}}{(2 M)^2} 
(M-M^*)-\frac{C_2^{\rm em}}{(2 M)^3} (M^* k) \right), \nn \\
A_{1/2}^{\frac{5}{2}^+} &=& \sqrt{2} e f_1
       \left( \frac{C_1^{\rm em}}{(2 M)^2} \frac{M}{M^*}(M-M^*)-
\frac{C_2^{\rm em}}{(2 M)^3} ( M^* k) \right),
\Ee
and for $N^*(D_{15})$
\Be
A_{3/2}^{\frac{5}{2}^-} &=& e f_1 (M-M^*)\left( 2 \frac{C_1^{\rm em}}
{(2 M)^2} - \frac{C_2^{\rm em}}{(2 M)^3} (M-M^* ) \right), \nn \\
A_{1/2}^{\frac{5}{2}^-} &=& e \sqrt{2} f_1 (M-M^*)
       \left( \frac{C_1^{\rm em}}{(2 M)^2} \frac{M}{M^*}-\frac{C_2^{\rm em}}{(2 M)^3} \frac{M- M^*}{2}\right),
\Ee
where
\be
f_1 = \frac{1}{2}\sqrt{\frac{k (E+M)}{10 M}},~ 
k=\frac{(M^{*2}-M^2)}{2 M^*}~. \nn
\ee

Although experimental values of the resonance coupling for the
hyperon resonances are not available, we can calculate the 
electromagnetic coupling
constants,
$g(Y^* Y \gamma)$, by using the nonrelativistic quark 
model-predicting resonance coupling amplitudes.
Similar to the case of the nucleon resonance, we can obtain 
the resonance coupling amplitudes
in terms of the helicity amplitude for radiative kaon capture.
The s-channel amplitude for the radiative kaon capture is related
to the u-channel amplitude for the kaon photoproduction by
crossing symmetry. Therefore, the helicity amplitude for the
hyperon resonance can be calculated by replacing the helicity
amplitude of kaon photoproduction,
${\cal M}_{-1/2,1-\lambda}^{\lambda_\gamma =1}$, in Eq. (\ref{helamp}),
with that of radiative kaon capture
where $\Gamma_K$ is the strong decay width of $Y^*$, i.e. $\Lambda
\rightarrow N +K^-$. The resulting resonance coupling amplitudes
of hyperon resonances $Y^*(L_{I 2J})$
are the identical with those of the nucleon resonances $N^*(L_{2 I 2J})$.

\def\prl#1{Phys.\ Rev.\ Lett.\ {\bf #1}}
\def\pr#1{Phys.\ Rev.\ {\bf #1}}
\def\prc#1{Phys.\ Rev.\ C {\bf #1}}
\def\prd#1{Phys.\ Rev.\ D {\bf #1}}
\def\np#1{Nucl.\ Phys.\ {\bf #1}}
\def\pl#1{Phys.\ Lett.\ {\bf #1}}
\def\prt#1{Phys.\ Reports.\ {\bf #1}}
\def\jpsj#1{J.\ Phys.\ Soc.\ Jpn.\ {\bf #1}}

\begin{table}
\caption[The particle list considered in our study.]
{List of the particles considered in our study.}
{
\setlength{\tabcolsep}{6.0 mm}
\begin{tabular}{|c|c|c|c|c|} \hline
particles   & $L_{2I(I)\,2J}$ & $J^P$            &mass (MeV)    
& $\Gamma$(MeV)   \\ \hline
$p$         &                 &  $\frac{1}{2}^+$ &938.28   &               \\ \hline
$\Lambda$   &                 &  $\frac{1}{2}^+$ &1115.6   &               \\ \hline
$\Sigma$    &                 &  $\frac{1}{2}^+$ &1192.46  &               \\ \hline
$K^+$       &                 & $0^-  $          &493.667  &               \\ \hline
$K^*$       &                 & $1^-  $          & 892     & 49.8          \\ \hline
$K_1$       &                 & $1^+ $           & 1270    & 90            \\ \hline
N1          & $P_{11}$        & $\frac{1}{2}^+$  & 1440    &  350          \\ \hline
N2          & $S_{11}$        & $\frac{1}{2}^- $ & 1535    & 150           \\ \hline
N3          & $S_{11}$        & $\frac{1}{2}^- $ & 1650    & 150           \\ \hline
N4          & $P_{11}$        & $\frac{1}{2}^+ $ & 1710    & 100           \\ \hline
N5          & $D_{13}$        & $\frac{3}{2}^-$  & 1520    & 120           \\ \hline
N6          & $D_{13}$        & $\frac{3}{2}^- $ & 1700    & 100           \\ \hline
N7          & $P_{13}$        & $\frac{3}{2}^+$  & 1720    & 150           \\ \hline
N8          & $D_{15}$        & $\frac{5}{2}^-$  & 1675    & 150           \\ \hline
N9          & $F_{15}$        & $\frac{5}{2}^+$  & 1680    & 130           \\ \hline
$\Lambda 1$ & $S_{01}$        & $\frac{1}{2}^-$  & 1405    & 50             \\ \hline
$\Lambda 2$ & $P_{01}$        & $\frac{1}{2}^+$  & 1600    & 150   \\ \hline
$\Lambda 3$ & $S_{01}$        & $\frac{1}{2}^-$  & 1670    & 35    \\ \hline
$\Lambda 4$ & $S_{01}$        & $\frac{1}{2}^-$  & 1800    & 300   \\ \hline
$\Lambda 5$ & $P_{01}$        & $\frac{1}{2}^+$  & 1810    & 150   \\ \hline
$\Sigma 1$  & $P_{11}$        & $\frac{1}{2}^+$  & 1660    & 100   \\ \hline
$\Sigma 2$  & $S_{11}$        & $\frac{1}{2}^-$  & 1750    &  90   \\ \hline
$\Lambda 6$ & $D_{03}$        & $\frac{3}{2}^-$  & 1520    & 15.6  \\ \hline
$\Lambda 7$ & $D_{03}$        & $\frac{3}{2}^-$  & 1690    & 60    \\ \hline
$\Sigma 3$  & $D_{13}$        & $\frac{3}{2}^-$  & 1670    & 60    \\ \hline
$\Lambda 8$ & $P_{03}$        & $\frac{3}{2}^+$  & 1890    & 100   
 \\ \hline
\end{tabular}
\label{nrlist1}
}
\end{table}

\begin{table}
\caption[Mixing angles for negative parity hyperon resonances]{
Mixing angles for negative parity hyperon resonances with $L_\sigma=P_M$.
All notations are the same as in Table \ref{tab:nrqA2}.}
\vskip 0.2cm
{\setlength{\tabcolsep}{1.7mm}
\begin{tabular}{|c|c|c|c|c|c|c|} \hline
   & & NRQM-predicting  & \multicolumn{4}{|c|}{mixing
  angles for $|{}^{2 S+1}XP_MJ^P\rangle$}\\ \cline{4-7}
 \raisebox{1.5ex}[0pt]{Resonances } &  \raisebox{1.5ex}[0pt]{ $J^P$}
 & mass (MeV)  & $|{}^21J^P\rangle$ & $|{}^2 8J^P\rangle$ & $|{}^4 8J^P\rangle$
 & $|{}^2 10J^P\rangle$ \\ \hline\hline
  $\Lambda 3$ &$\frac{1}{2}^-$& 1650 & $-0.39$ & 0.75   & $-0.58$  &         \\
  \hline
  $\Lambda 4$ &$\frac{1}{2}^-$& 1800 & -0.18 & 0.5    & 0.85   &         \\ \hline
  $\Sigma 2 $ &$\frac{1}{2}^-$& 1810 &     & $-0.33$  & $-0.21$  & 0.92    \\
  \hline
  $\Lambda 6$ &$\frac{3}{2}^-$& 1490 & 0.91  & 0.4    & 0.01   &         \\ \hline
  $\Lambda 7$ &$\frac{3}{2}^-$& 1690 & $-0.4$  & 0.91   & 0.12   &         \\ \hline
\end{tabular}
\label{tab:nrqA1}
}
\end{table}

\begin{table}
\caption[Photon coupling amplitudes for the states in 1st column
to $\Lambda \gamma$.]{Photon coupling amplitude for the states in 1st column
to $\Lambda \gamma$. $K=q^2/6\alpha^2$ where $q$ is the photon
momentum and $x=m_u/m_s$. The full photon coupling amplitudes are
obtained by multiplying entries in the table by $i \sqrt{4 \pi}
\mu_p \alpha e^{-K}$. The parameters are given as follows\cite{Isgu77}: $\alpha=0.41$ GeV,
$\mu_p$=0.13 GeV, and $x=0.6$. In this Table we use the notation 
$|{}^{2 S+1}XL_\sigma J^p\rangle$.
Here X is the SU(3) multiplicity, $S$, $L$, $P$, and $J$ are the total spin, total orbital angular momentum,
parity, and total angular momentum, and $\sigma$ is the permutational 
symmetry of the SU(6) state.}
\vskip 0.2cm
{\setlength{\tabcolsep}{7.0mm}
\begin{tabular}{|l||c|c|}\hline
   & \multicolumn{2}{|c|}{photon coupling amplitudes for decay}\\
    $ | {}^{2 S +1}X L_\sigma J^P\rangle$ &
    \multicolumn{2}{|c|}{$|{}^{2 S +1}X L_\sigma J^P\rangle\rightarrow \Lambda +\gamma$} \\ \cline{2-3}
 & $A_{3/2}$  & $A_{1/2}$  \\ \hline
 $|\Lambda^21 P_M \frac{3}{2}^- \rangle$  &  $-\frac{1}{\sqrt{6}}\left(\frac{2 x +1}{3}\right)$   &
 $-\frac{\sqrt{2}}{\sqrt{6}}\left(\frac{2 x +1}{3} (1-6 K)\right)$
 \\ \hline
$|\Lambda^28 P_M \frac{3}{2}^- \rangle$   &   $-\frac{1}{\sqrt{6}}\left(\frac{2 x +1}{3}\right)$  &
$-\frac{1}{\sqrt{6}}\left(\frac{2 x +1}{3} -(2 x-1) 2 K \right)$
\\ \hline
$|\Lambda^48 P_M \frac{3}{2}^- \rangle$   &  $\frac{\sqrt{15}}{5} K$  & $\frac{1}{3 \sqrt{5}} K$
\\ \hline\hline
 $|\Lambda^21 P_M \frac{1}{2}^- \rangle$ &not  & $ -\frac{1+3 K}{3}\left(\frac{2 x+1}{3}\right) $
 \\ \hline
 $|\Lambda^28 P_M \frac{1}{2}^- \rangle$ &not  & $ -\frac{1}{3}\left(\frac{2 x+1}{3}+(2 x-1) K \right) $
  \\  \hline
 $|\Lambda^48 P_M \frac{1}{2}^- \rangle$ &not  & $ \frac{K}{3} $    \\ \hline
 $|\Sigma^28 P_M \frac{1}{2}^- \rangle$ &not  & $ \frac{1}{\sqrt{3}}\left( 1+\frac{1}{6} K\right) $
 \\ \hline
 $|\Sigma^48 P_M \frac{1}{2}^- \rangle$ &not  & $ -\frac{1}{6 \sqrt{3}} K $
  \\  \hline
 $|\Sigma^2 10 P_M \frac{1}{2}^- \rangle$ &not  & $ -\frac{1}{\sqrt{3}}\left(1-\frac{1}{6} K\right) $    \\ \hline
\end{tabular}
\label{tab:nrqA2}
}
\end{table}

\begin{table}
\caption[The obtained photon coupling amplitudes for odd parity hyperon resonances
within the NRQM framework.]{The obtained photon coupling amplitudes for odd parity hyperon resonances
within the NRQM framework.}
\vskip 0.2cm
{\setlength{\tabcolsep}{5.7mm}
\begin{tabular}{|l||c|c|c|}\hline
   &   & \multicolumn{2}{|c|}{The photon coupling amplitudes }\\   \cline{3-4}
\raisebox{1.5ex}[0pt]{Hyperon resonances}&\raisebox{1.5ex}[0pt]{$J^P$} & $A_{3/2}$  & $A_{1/2}$  \\ \hline
  $\Lambda 3$&$\frac{1}{2}^-$ &  &  -0.080        \\
  \hline
  $\Lambda 4$ &$\frac{1}{2}^-$ &  &  0.038       \\ \hline
  $\Sigma 2 $ &$\frac{1}{2}^-$ &     & -0.70   \\
  \hline
  $\Lambda 6$  &$\frac{3}{2}^-$ & -0.082 & -0.026         \\ \hline
  $\Lambda 7$ &$\frac{3}{2}^-$  & -0.020  & -0.024        \\ \hline

\end{tabular}
\label{tab:nrqA3}
}
\end{table}

\begin{table}
{\renewcommand\baselinestretch{0.9}
\caption[Branching ratios and coupling constants
of the particles included in our study.]{Branching
ratios and coupling constants of the particles included in our study. The strong and radiative decay branching ratios are  respectively
shown in column 2 and 3, and the experimental values for the
resonance couplings of photon are given in forth column \cite{PDG98}.
In the last two columns, $g_s$ and $g_{E.M}$ respectively represent a strong and an electromagnetic
coupling constants where  $g_s=g_{KB_1 B_2}/\sqrt{4 \pi}$, $g_s=f_{KB_1 B_2}/m_K$,
 and $g_s=f_{KB_1 B_2}/m_K^2$ for
         $J=\frac{1}{2}$, $J=\frac{3}{2}$, and $J=\frac{5}{2}$ particles, respectively.
The values in parentheses are NRQM-predicting ones \cite{Isgu77,Juri83}.}
\small
{\setlength{\tabcolsep}{1.0 mm}
\begin{tabular}{|c|c|c|c|c|c|} \hline
particle& $\frac{\Gamma_{{}_{N_R ->K\Lambda}}}{\Gamma} (\%)$ &
$\frac{\Gamma_{{}_{N_R ->\gamma N}}}{\Gamma}(\%)$& ${{\rm Photon} \atop {\rm coupling}}$ (A) &$g_s$ &$g_{E.M}$ \\ \hline\hline
p & * & * & * & - & - \\ \hline
$\Lambda$ & * & * & * & -& - \\ \hline
$\Sigma$ &  * & * & * &-&- \\ \hline
$K^+ $ & * & * & * &-&-\\ \hline
$K^*  $ & * & $0.101 \pm 0.009 $& * & - & - \\ \hline
$K_1 $  & *& * &*  & - & -\\ \hline
N1  &  *& 5 - 10 & $A_{1/2}=-0.072 \pm  0.009$ &-& 0.57\\ \hline
N2 &  *  & 0.45 - 0.53 & $A_{1/2}= 0.068 \pm 0.010$ &- &0.89\\ \hline
N3 & 3 -11 & 0.10 - 0.18 & $A_{1/2}= 0.052 \pm 0.017$&0.21&0.38  \\ \hline
N4  & 5-25 &  *  & $A_{1/2}= -0.006 \pm 0.027$&1.72&(0.032) \\ \hline
   & & &    $ A_{1/2}= -0.022 \pm 0.008$ & &  \\ \cline{4-4}
\raisebox{1.5ex}[0pt]{N5} &    \raisebox{1.5ex}[0pt]{ * } &
       \raisebox{1.5ex}[0pt]{0.45 - 0.53} &$A_{3/2}=+0.163 \pm 0.007$ &
\raisebox{1.5ex}[0pt]{-}&\raisebox{1.5ex}[0pt]{-}\\ \hline
  & & & $ A_{1/2}=-0.017 \pm 0.012$  & & \\ \cline{4-4}
\raisebox{1.5ex}[0pt]{N6 }& \raisebox{1.5ex}[0pt]{ $< 3$ }& \raisebox{1.5ex}[0pt]{$ \sim 0.01$} &
$A_{3/2}=  +0.002 \pm 0.02$ &\raisebox{1.5ex}[0pt]{14.2}&\raisebox{1.5ex}[0pt]{-}\\ \hline
  & & & $A_{1/2}=+0.027 \pm 0.024$  & & \\ \cline{4-4}
\raisebox{1.5ex}[0pt]{N7}  & \raisebox{1.5ex}[0pt]{1 -15} &
     \raisebox{1.5ex}[0pt]{0.01 -0.06} & $ A_{3/2}=-0.026 \pm 0.010 $ &
     \raisebox{1.5ex}[0pt]{3.89}&
     \raisebox{1.5ex}[0pt]{-} \\ \hline
  & & & $A_{1/2}=+0.019 \pm 0.008$  & & \\ \cline{4-4}
\raisebox{1.5ex}[0pt]{N8}  & \raisebox{1.5ex}[0pt]{$<1$} &
     \raisebox{1.5ex}[0pt]{0.004 - 0.023} & $ A_{3/2}=+0.015 \pm 0.009 $ &
                           \raisebox{1.5ex}[0pt]{-}&
     \raisebox{1.5ex}[0pt]{-} \\ \hline
   & & & $A_{1/2}=-0.015 \pm 0.006$  & & \\ \cline{4-4}
\raisebox{1.5ex}[0pt]{N9}  & \raisebox{1.5ex}[0pt]{$*$} &
     \raisebox{1.5ex}[0pt]{0.21 - 0.32} & $ A_{3/2}=+0.133 \pm 0.012 $ &
                           \raisebox{1.5ex}[0pt]{-}&
     \raisebox{1.5ex}[0pt]{-} \\ \hline
$\Lambda 1$ &  *  &  * & * &-&-\\ \hline
$\Lambda 2$ & 15 - 30 &  *  & *&1.14&- \\ \hline
$\Lambda 3$ & 15 - 25  & * & * &0.084&(0.094)\\ \hline
$\Lambda 4$ & 25 - 40 & * & * &0.2&(0.117)\\ \hline
$\Lambda 5$ & 20 - 50 & * & * &0.79 &- \\ \hline
$\Sigma 1$ & 10 - 30 & * & * & 0.70& - \\ \hline
$\Sigma 2$& 10 - 40 & * & * & 0.14& (0.102) \\ \hline
  & & & $(A_{1/2}=-0.0256)$  & & $(-0.993)$ \\ \cline{4-4} \cline{6-6}
     \raisebox{1.5ex}[0pt]{$\Lambda 6$ } & \raisebox{1.5ex}[0pt]{$45\pm1$} &
     \raisebox{1.5ex}[0pt]{$0.8\pm 0.2$} & $ (A_{3/2}=-0.0821 )$ &
                           \raisebox{1.5ex}[0pt]{21.1}&
     $(-0.256)$ \\ \hline
  & & & $(A_{1/2}=-0.0244)$  & & (0.422) \\ \cline{4-4}\cline{6-6}
     \raisebox{1.5ex}[0pt]{$\Lambda 7$ }  & \raisebox{1.5ex}[0pt]{$20 - 30$} &
     \raisebox{1.5ex}[0pt]{$*$} & $ (A_{3/2}=-0.02034 )$ &
                           \raisebox{1.5ex}[0pt]{7.87}&
     $(0.851)$ \\ \hline
   & & & $*$  & & - \\ \cline{4-4}\cline{6-6}
     \raisebox{1.5ex}[0pt]{$\Sigma 3$} & \raisebox{1.5ex}[0pt]{$7 - 13$} &
     \raisebox{1.5ex}[0pt]{$*$} &$*$ &
                           \raisebox{1.5ex}[0pt]{5.51}&
     - \\ \hline
$\Lambda 8$  & 20-35 & *& * &1.49 & - \\ \hline
\end{tabular}
\label{nrlist2}
}
$*$ no experimental data.
}
\end{table}
\begin{table}
\caption[The sensitivity of $\chi^2$ with respect to the leading coupling constants and
the odd parity hyperon resonances]{
The sensitivity  of $\chi^2$ with respect to the leading coupling constants and
the odd parity hyperon resonances. Others unlisted in
this table but listed in Table \ref{tab:result2} are allowed to
vary in the explained ranges in this Section.
Of the  given values, $g_{KN\Lambda}$ and $g_{KN\Sigma}$ are  the given
values in Eq. (\ref{eqgfix}) and
 the others are calculated by NRQM.
'x' means that  the parameter is permitted to vary in the permitted range
explained in Section \ref{secfit} and the blank
represents that the corresponding particles are excluded in the calculation.}
 \setlength{\tabcolsep}{3.0 mm}
\begin{tabular}{|c|c|c|c|c|c|c||c||}\hline
                                 & A  &  B     &  C   &  D  &   E    &  F  &   G    \\ \hline\hline
  $g_{KN\Lambda}/\sqrt{4 \pi}$   & x  &  x     & -3.74&-3.74&  -3.74 &-3.74& -3.740 \\ \hline
  $g_{KN\Sigma}/\sqrt{4 \pi} $   & x  &  x     & 1.09 &1.09 &   1.09 & 1.09&  1.090 \\ \hline
  $ G_{\Lambda 3}/\sqrt{4 \pi} $ & x  &  x     & x    &  x  &  0.007 &  x  &  0.008 \\ \hline
  $ G_{\Lambda 4}/\sqrt{4 \pi} $ & x  &  x     & x    &  x  &  0.023 &  x  &  0.023 \\ \hline
  $ G_{\Sigma 2}/\sqrt{4 \pi} $  & x  &  x     & x    &  x  &  0.014 &  x  &  0.014 \\ \hline
  $ G^1_{\Lambda 6} $            & x  & -20.94 & x    &-20.9&  -20.9 &     &        \\ \hline
  $  G^2_{\Lambda 6} $           & x  & -5.402 & x    &-5.40&  -5.40 &     &        \\ \hline
  $  G^1_{\Lambda 7} $           & x  &  3.325 & x    & 3.32&   3.32 &     &        \\ \hline
  $  G^2_{\Lambda 7} $           & x  &  6.700 & x    & 6.70&   6.70 &     &        \\ \hline
  $  G^1_{\Sigma 3} $            & x  &  x     & x    & x   &   x    &     &        \\ \hline
  $ G^2_{\Sigma 3} $             & x  &  x     & x    & x   &   x    &     &        \\ \hline\hline
  $ \chi^2/N         $           &0.97&  1.0   & 1.0 & 1.0 &   1.0  & 1.0 &  1 .0  \\ \hline
\end{tabular}
\label{tab:result1}
\end{table}
\begin{table}
\def\baselinestretch{1.15}
\caption[Parameters of our models]{Parameters of our models:
Parameters in the 2nd and the 3rd columns
are obtained by using
the Haberzettl's gauge invariance prescription. Those of 
the Ohta's prescription are
 in 4th and 5th columns  and  the cases without strong form factors are 
in the last two columns.
PV(PS) means that the
coupling constants in the column are obtained by using
pseudovector(pseudoscalar) coupling scheme. The coupling constants of spin-1 and spin-$\frac{1}{2}$ are
defined in Section \ref{Clag} and those of spin-$\frac{3}{2}$ and spin-$\frac{5}{2}$ are defined in \cite{hanphd}. 
The parameters, $\Lambda_J$'s in last four lines
are cutoff parameters for hadronic form factors corresponding
spin=$J$ particles, respectively.
}
{\small
 \setlength{\tabcolsep}{0.5 mm}
\begin{tabular}{ccccccc} \hline
      parameters                   & Habe(PV) &Habe(PS) &Ohta(PV)   & Ohta(PS)  &Noform(PS) &Noform(PV)\\ \hline\hline
$g_{KN\Lambda}/\sqrt{4 \pi}$ & -3.74    &   -3.74 &   -3.74   &   -3.74   &  -3.74   & -3.74    \\ \hline
$g_{KN\Sigma}/\sqrt{4 \pi} $ & 1.09     &    1.09 &    1.09   &    1.09   &   1.09   &  1.09    \\ \hline
$G^{K^*}_V/4\pi            $ & -0.16    &  -0.22  &   0.37    &   0.37    &  -0.34   &-0.19     \\ \hline
$G^{K^*}_T/4\pi            $ & -0.83    &  -0.83  &  -0.83    &  -0.83    &  -0.03   & -0.33    \\ \hline
$G^{K_1}_V/4\pi            $ & -0.0042  &  -0.01  &   0.084   &  0.082    &    0.26  &   0.54   \\ \hline
$G^{K_1}_T/4\pi            $ & -0.305   &  -0.39  &   -3.41   &  -3.48    &   -0.56  & -0.71    \\ \hline
$G_{N1}/\sqrt{4\pi} $        &  -3.78   &   -3.11 &   -4.11   &  -3.09    &    0.61  &   4.40   \\ \hline
$G_{N2}/\sqrt{4\pi} $        &  0.104   &   0.19  &   0.12    &  0.18     &    0.03  &  -0.23   \\ \hline
$G_{N3}/\sqrt{4\pi} $        &  -0.022  &  -0.015 &   0.0057  &  0.014    &    0.02  &   0.05   \\ \hline
$G_{N4}/\sqrt{4\pi} $        &   0.179  &   0.22  &   0.0853  &  0.091    &   -0.05  & -0.13    \\ \hline
$G_{\Lambda 1}/\sqrt{4\pi} $ &   -1.51  &   -1.41 &  -0.26    & -0.27     &   -0.31  & -0.38    \\ \hline
$G_{\Lambda 3}/\sqrt{4\pi} $ &   0.008  &   0.008 &   0.008   &  0.008    &   0.008   &  0.008   \\ \hline
$G_{\Lambda 4}/\sqrt{4\pi} $ &   0.023  &   0.023 &   0.023   &  0.023    &   0.023   &  0.023   \\ \hline
$G_{\Sigma 2}/\sqrt{4\pi} $  &   0.014  &   0.014 &   0.014   &  0.014    &   0.014   &  0.014   \\ \hline
$G_{N6}^1 $                  & 32.154   &    23.8 &    23.62  &   20.84   &   3.586   &  7.634   \\ \hline
$G_{N6}^2 $                  & 42.965   &    37.9 &    30.88  &   29.78   &   1.484   & -4.201   \\ \hline
$G_{N7}^1 $                  &  -0.076  &  0.36   &   0.70    &   0.66    &  -0.353   & -0.468   \\ \hline
$G_{N7}^2 $                  &    1.00  &  1.33   &    2.17   &   2.42    &  -0.927   &  1.585    \\ \hline
$G_{N8}^1 $                  &   -40.85 & -35.27  &   -36.49  &  -38.92   &   9.699   &  2.297    \\ \hline
$G_{N8}^2 $                  &    34.92 &  10.95  &    18.11  &   4.69    & -28.114   & -6.864    \\ \hline
$\Lambda_{1}^2  $            &   0.50   &  0.50   &   0.80    &   0.80    &           &            \\ \hline
$\Lambda_{1/2}^2$            &   0.60   &  0.88   &   0.53    &   0.76    &           &            \\ \hline
$\Lambda_{3/2}^2$            &   0.79   &  0.89   &    1.13   &   1.15    &           &            \\ \hline
$\Lambda_{5/2}^2$            &   0.75   &  0.79   &   0.75    &   0.79    &           &           \\ \hline\hline
$\chi^2/N$                   &   1.0    &  1.04   &  1.24     &  1.26     & 1.3       &
2.72  \\ \hline
\end{tabular}
\label{tab:result2}
}
\end{table}

\begin{table}
\caption[The partial $\chi^2$ per data]{The partial $\chi^2$ per data:
$\sigma_{\rm tot}$, $d\sigma/d\Omega$, P, and  T are the $\chi^2$ values
of the total and differential cross sections,
$\Lambda$-polarization asymmetry and  a target polarization asymmetry,
respectively. The last column is a total $\chi^2$ per particle.}
\vskip 0.2 cm
 \setlength{\tabcolsep}{5.6 mm}
\begin{tabular}{|c|c|c|c|c|c|}\hline
$\chi^2 \rightarrow$  &   $\sigma_{\rm tot}$ & $d\sigma/d\Omega$ & P & T &$ \chi^2_{\rm tot}$ \\ \hline\hline
Habe(PV) &  0.975 &  0.957  & 0.993   & 4.36 & 1.0    \\ \hline
Habe(PS) &  1.01& 0.99  &  1.11   &  3.16  &   1.03 \\ \hline
Ohta(PV) &  1.22 &   1.07 &  1.055  &  8.13  &   1.24  \\ \hline
Ohta(PS) &  1.27 &   1.08 &  1.008  &  3.97  &   1.26  \\ \hline
Noform(PV) &  5.66 &   1.7 &  4.19  &  9.39  &   2.72  \\ \hline
Noform(PS) &  2.3 &   1.2 &  1.17  &  0.3  &   1.3  \\ \hline
\end{tabular}
\label{tab:result3}
\end{table}

\begin{table}
\caption[Branching ratio of the $K^- +p \rightarrow \gamma
+\Lambda$]{Branching ratio of the $K^- +p \rightarrow \gamma
+\Lambda$ for each model. Experimental data are taken from Ref. \cite{Whit89}.
Unit of ($\times 10^{-3}$)}
\vskip 0.2 cm
\begin{tabular}{|c|c|c|c|c|c|c|}\hline
 \multicolumn{2}{|c|}{Haberzettl} & \multicolumn{2}{c|}{Ohta} & \multicolumn{2}{c|}{Without form factors}
 &  \\ \cline{1-6}
 PV & PS&PV&PS&PV&PS& \raisebox{0.75em}[0pt]{Experimental value} \\  \hline \hline
0.87 & 0.94&1.35 &1.51 & 1.82 & 1.16  &$0.86 \pm 0.07\pm0.09$ \\
\hline
\end{tabular}
\label{tab:result4}
\end{table}

\end{document}